\documentclass[%
 aip,
 amsmath,amssymb,
 reprint,%
floatfix
]{revtex4-1}

\usepackage{graphicx}
\usepackage{dcolumn}
\usepackage{bm}

\usepackage[utf8]{inputenc}
\usepackage[T1]{fontenc}
\usepackage{mathptmx}
\usepackage{etoolbox}
\usepackage[ruled,vlined]{algorithm2e} 
\usepackage{physics}

\makeatletter
\def\@email#1#2{%
 \endgroup
 \patchcmd{\titleblock@produce}
  {\frontmatter@RRAPformat}
  {\frontmatter@RRAPformat{\produce@RRAP{*#1\href{mailto:#2}{#2}}}\frontmatter@RRAPformat}
  {}{}
}%
\makeatother

\usepackage{xcolor}

\begin{document}

\preprint{AIP/123-QED}

\title[Lanczos algorithm for SOS calculations of SSCCs]{
A Lanczos-based algorithm for sum-over-states calculations of NMR spin--spin coupling constants at the RPA level of theory: The Fermi-contact term
}

\author{Sarah L. V. Zahn}
\affiliation{Department of Chemistry, University of Copenhagen, Universitetsparken 5, 2100 Copenhagen, Denmark}
\author{Luna Zamok}
\affiliation{Department of Chemistry, University of Copenhagen, Universitetsparken 5, 2100 Copenhagen, Denmark}
\author{Sonia Coriani}
\affiliation{Department of Chemistry, Technical University of Denmark, Kemitorvet Bldg 207, 2800 Kongens Lyngby, Denmark}
\author{Stephan P. A. Sauer}
\email{sauer@chem.ku.dk}
\affiliation{Department of Chemistry, University of Copenhagen, Universitetsparken 5, 2100 Copenhagen, Denmark}

\date{\today}

\begin{abstract}
The analysis of nuclear magnetic resonance parameters, such as the indirect nuclear spin–spin coupling constants, in terms of contributions from localised molecular orbitals is a commonly used approach for gaining a deeper understanding of experimentally observed trends in these parameters.
In the vast majority of these studies, contributions from pairs of one occupied and one virtual orbital are calculated and analyzed. 
Analyses in terms of two pairs of an occupied and a virtual orbital, that would allow for the study of coupling pathways, are much more seldom, as they require calculating the coupling constants as a sum over all excited states. 
Previous studies have shown that, for the often dominating Fermi-contact contribution to the coupling constants, more or less all excited states have to be calculated when employing a Davidson algorithm, because the most high-lying excited states can also make a significant contribution to the Fermi-contact term.
In this study we investigated therefore, whether by employing a Lanczos algorithm one can obtain converged values of the Fermi-contact contribution to the indirect nuclear spin-spin coupling constants already with a significantly smaller percentage of the total number of excited states included in the sum-over-states expression. 
To this purpose we have extended the recent implementation of a Lanczos algorithm for the RPA/TDHF or TDDFT eigenvalue problem in the Dalton program (L. Zamok \textit{et al.} J. Chem. Phys. 156, 014102 (2022)).
The new procedure was tested on 17 molecules containing first, second and third row atoms.
We find that, for most coupling constants, less than 50\% of the excited pseudo states are necessary for converging the Fermi-contact term with an error of less than 0.5 Hz.
For the few exceptions, typically for molecules with third-row atoms, around 60\% were necessary.
\end{abstract}

\maketitle

\section{Introduction}
The analysis of NMR spin-spin coupling constants (SSCC) in terms of contributions from individual pairs of localized occupied and virtual orbitals, such as bonding and anti-bonding orbitals, has a long and well-established tradition.\cite{JCE-84-156-2007-Autschbach} 
Many important effects, like for instance stereoelectronic or stereochemical effects, anomeric effects, electron lone pair effects or hyperconjugative interactions on SSCCs, have been studied in this way.

Three general approaches are commonly employed for such analyses. 
The earliest methodology relies on Ramsey’s sum‑over‑states (SOS) expressions for the SSCCs\cite{nmr530715r} 
in which the excitation energies are approximated using the eigenvalues of 
electronic 
Hessians or principal propagators.\cite{spasB8} 
Within this framework, the associated transition moments are derived from property integrals in the molecular orbital basis in combination with the eigenvectors of the principal propagator.
%
Transforming both quantities to the basis of localized molecular orbitals leads to expressions for the coupling constants as a sum over contributions from simultaneously two localized occupied and two localized unoccupied orbitals, which allows for the analysis of coupling pathways through a molecule.
This is called the Contributions from Localized Orbitals within the Polarization Propagator Approach and Inner Projections of the Polarization Propagator (IPPP-CLOPPA), which originally used 
electronic 
Hessians calculated with semi-empirical methods.\cite{nmr83-ijqc23-1033,CPC-39-409-1986-Engelmann,IJQC-S20-585-1986-Azua,nmr90rgac,nmr90drgc} 
In more recent years, ab initio methods at the level of Hartree-Fock \cite{IJQC-S20-603-1986-Scuseria,JCC-19-181-1998-Giribet,JMS-T-433-141-1998-Azua} or density functional theory \cite{spas077,spas084,spas153,JCP-126-174103-2007-Autschbach} have also been employed in such calculations and analyses.

In the second approach, the coupling constants are expressed as a sum over contributions involving only a single occupied–virtual orbital pair at a time. As a consequence, some of the information retained in the CLOPPA method is lost.
On the other hand, this approach does not require the explicit calculation of individual excited states of a molecule. Instead, it is based on an analysis of the solution vectors of an approximate linear‑response function, or on the first‑order perturbed orbitals obtained from coupled‑perturbed density function theory (DFT) or Hartree–Fock calculations, subsequently transformed to localized orbitals. 
This methodology was probably pioneered by Contreras and co-workers\cite{Adcock1999,Della2000,nmr01-jpca105-5298,Barone2001a,MRC-39-600-2001-Barone,Sosa2002,MP-100-705-2002-Zaccari,IJMS-4-93-2003-Zaccari,IJQC-110-532-2010-Contreras} and has frequently been employed by others\cite{MRC-50-665-2012-Rusakov,MRC-50-653-2012-Rusakov,Vega-2013,JMM-20-2225-2014-Vega,ACS-Omega-4-1494-2019-Nepel,JPCA-123-10072-2019-Zeoly,MRC-online-2022-Soares} in combination with Weinberg's natural bond orbitals.\cite{JACS-123-12026-2001-Wilkens} 
More recently, it was also extended to the level of the second order polarization propagator approximation (SOPPA)\cite{Zarycz2012a,Zarycz2012,spas145} and to relativistic DFT calculations using the zeroth-order regular approximation (ZORA).\cite{Autschbach2007g,glentOM2021} 

Several years ago, Cremer and co-workers\cite{nmr03-jpca107-7043, nmr04-jcp120-9952, nmr04-cpl383-332,nmr04-cpl387-415,Grafenstein2004d,nmr07-pccp9-2791} developed  a third approach at the 
coupled-perturbed DFT level, where the contributions from particular orbitals were obtained indirectly by removing these orbitals in the calculations.

The advantage of the original CLOPPA or SOS approach, in which one obtains contributions from two pairs of occupied and virtual orbitals, is that it emphasizes the simultaneous perturbation of the electrons by both nuclei and thus the coupling pathways. 
Furthermore, it also allows the contributions to the coupling constant from different excited states to be examined.\cite{spas153} 
At the same time, this feature also constitutes the main limitation of the approach, since previous experience has shown that all excited states, including pseudo‑states representing the continuum, must be included.\cite{jod14} For molecules larger than those studied previously,\cite{spas077,spas084} this requirement becomes prohibitively expensive.

In a recent study, Zarycz et al. \cite{spas153} investigated therefore whether one could truncate the number of excited states or pseudo-states used in the calculation of SSCCs. 
Their study was carried out at the DFT level using the B3LYP functional and an adaption of the Davidson algorithm,\cite{DavidsonJCoP1975} which preserves the paired structure of the time-dependent DFT eigenvalue problem.\cite{OlsenJCoP1988} 
This algorithm will, when asked for $N$ eigenvalues of a matrix with dimension $N_{max} > N$, produce the $N$ energies with the lowest absolute values converged to a predefined accuracy. 
The authors only considered the Fermi-contact term, since it was the dominant term for their test molecules. 
They showed that, when using the Davidson algorithm, the one‑bond coupling constants can fluctuate until all excited states are included in the summation. This indicates that even pseudo‑states with  
the highest
excitation energies play a significant role and must be included in the summation.
However, they also found that the different types of couplings behaved differently. 
For example, the geminal coupling constants, $^2$J(HH), quickly converged, with occasional big spikes stemming from calculations where not all degenerate states were simultaneously included. 
Nevertheless, the question arises whether one could achieve a faster convergences of the SOS expressions for the coupling constants with an algorithm that yields converged or approximate values for the lowest and highest excitation energies at the same time as, e.g., the Lanczos algorithm.\cite{lanczosJNBS1950}


In 1950 Cornelius Lanczos presented what he called ``An Iteration Method for the Solution of the Eigenvalue Problem of Linear Differential and Integral Operators'', which later became known as the Lanczos algorithm.\cite{lanczosJNBS1950}
When asked for $N$ eigenvalues of a matrix with a dimension $N_{max} > N$,  the Lanczos algorithm will return approximations to both the largest and smallest eigenvalues, 
in contrast to the (converged) $N$ lowest eigenvalues the Davidson algorithm yields.\cite{DavidsonJCoP1975} Expressed in a different way, the Lanzcos algorithm converges the spectrum of eigenvalues from the top and the bottom, and only after $N_{max}$ iterations will all eigenvalues be fully converged. This implies that, for $N < N_{max}$, the $N$ eigenvalues in the Lanczos algorithm are only approximations to the eigenvalues but cover a larger part of the eigenvalue spectrum than in the Davidson algorithm,\cite{DavidsonJCoP1975} 
where they would be converged values of the $N$ lowest eigenvalues.
Throughout the years, several adaptations of the Lanczos algorithm have been implemented.\cite{johnsonCPC1999,chernyakJCP2000,tsiperJPB2001,roccaJCP2008,tretiakJCP2009,hansenJCP2010,corianiJCTC2012,ZamokJCP2021}
Two studies, in particular, are of interest here, since they explicitly used the algorithm to calculate SOS properties. 
The first study was presented by Johnson et al.\cite{johnsonCPC1999} in 1999. 
They implemented a version for the time-dependent Hartree-Fock (TD-HF) or random phase approximation (RPA) eigenvalue problem that took advantage of the paired structure of the RPA eigenvalue problem and preserved its form throughout the algorithm.\cite{johnsonCPC1999}
In the second study by Zamok et al.,\cite{ZamokJCP2021} Johnson's algorithm was extended to the calculation of sum-over-states properties and implemented in the open-source quantum chemistry program Dalton.\cite{daltonpaper,spas191} 
The authors used the algorithm to calculate mean excitation energies,\cite{spas185} and found that they could accurately reproduce their values by only including 10-30$\%$ of the excited states.
Their version of the Lanczos algorithm, as it preserves the paired structure of the RPA eigenvalue problem with its positive and identical but negative eigenvalues, converges the largest positive and smallest negative eigenvalues first . 
For SOS properties like the mean excitation energies or the SSCCs, however, only the positive eigenvalues are needed. This implies that, with a Lanczos algorithm, one starts with the excited states with highest energy, and at each iteration one obtains more accurate values of these energies and at the same time adds a state with lower energy, in total contrast to the Davidson algorithm.\cite{DavidsonJCoP1975} 
For the mean excitation energies, whose values are very much dominated by the pseudo-states representing the continuum, this  turned out to be optimal.

Based on this success,\cite{ZamokJCP2021} 
we asked ourselves whether by using this Lanczos algorithm in SOS calculations of spin-spin coupling constants one could obtain converged results with a significantly smaller number of excited states than necessary when using the Davidson algorithm.\cite{DavidsonJCoP1975} 
This would then allow us to carry out CLOPPA analyses of coupling constants also for larger molecules than presently possible.\cite{spas077,spas084}

In the following, we first recapitulate 
the most important details about calculations of SSCCs and on the Lanczos algorithm. Afterwards, we present the necessary changes to the Lanczos algorithm, previously implemented by Zamok et al.,\cite{ZamokJCP2021} for calculations of SSCCs. Finally, we investigate with calculations on 17 molecules, whether with this algorithm it is possible to truncate the number of states in the SOS calculation of SSCCs. 

\section{Theory}
\subsection{The NMR spin--spin coupling constant as a sum-over-states property}
The NMR coupling constant $J$ for two nuclei $N$ and $M$ consists of four terms:\cite{ramseyPR1953} the Fermi-contact term (FC), the spin-dipolar term (SD), the paramagnetic spin-orbit term (PSO), and the diamagnetic spin-orbit term (DSO) 
\begin{align}\label{eq:J}
    J(N,M) = J^\textrm{FC}(N,M) + J^\textrm{SD}(N,M) + J^\textrm{PSO}(N,M) + J^\textrm{DSO}(N,M) .
\end{align}
Three of these can be calculated as static linear response functions, with respect to the three perturbation operators $\hat{\vec O}^\textrm{FC}$, $\hat{\vec O}^\textrm{SD}$ and $\hat{\vec O}^\textrm{PSO}$; the DSO term, on the other hand, is usually calculated as an expectation value of the operator $\hat{\mathbf{O}}^\textrm{DSO}$, although a reformulation as linear response function also exists,\cite{spas007,nmr12-jcp137-074108}
\begin{align}
\label{eq:sscc_lr}
    J(N,M) &= 
    \frac{1}{3}
    \frac{\gamma_N\gamma_M}{h} \displaystyle
    \sum_{\alpha=x,y,z} 
    \left \{\langle\langle\hat{O}^\textrm{FC}_{N\alpha}+
\hat{O}^\textrm{SD}_{N\alpha};\hat{O}^\textrm{FC}_{M\alpha}+
\hat{O}^\textrm{SD}_{M\alpha}\rangle\rangle_{\omega=0} \right . \\ 
    &+ \left . \langle\langle\hat{O}^\textrm{PSO}_{N\alpha};\hat{O}^\textrm{PSO}_{M\alpha}\rangle\rangle_{\omega=0} + \mel{\Psi_0}{\hat{O}^\textrm{DSO}_{NM,\alpha\alpha}}{\Psi_0}
    \right \}~. \nonumber
\end{align}
The four perturbation operators are defined as
\begin{align}\label{eq:o_fc}
    \hat{O}^\textrm{FC}_{N\alpha} = \frac{\textit{g}_e e \mu_0}{3m_e} \displaystyle\sum_{i}^{N_e}s_{i\alpha}\delta(r_{iN}) ,
\end{align}
\begin{align}\label{eq:o_sd}
    \hat{O}^\textrm{SD}_{N\alpha} = 
    \frac{\textit{g}_e e \mu_0}{8\pi m_e}
    \displaystyle\sum_{i}^{N_e}\frac{3(\vec s_i\cdot \vec r_{iN})r_{iN,\alpha}-r^2_{iN}s_{i,\alpha}}{r^5_{iN}} ,
\end{align}
\begin{align}\label{eq:o_pso}
    \hat{O}^\textrm{PSO}_{N\alpha} = 
    \frac{e \mu_0}{4\pi m_e}\displaystyle\sum_{i}^{N_e}
    \frac{l_{iN,\alpha}}{r^3_{iN}} ,
\end{align}
\begin{align}\label{eq:o_dso}
    \hat{O}^\textrm{DSO}_{NM,\alpha\alpha} = 
    \left(\frac{\mu_0}{4\pi}\right)^2 \frac{e^2}{2 m_e}
    \displaystyle\sum_{i}^{N_e} 
    \frac{\vec r_{iM}\cdot \vec r_{iN} - r_{iM,\alpha}r_{iN,\alpha}}{r^3_{iM}r^3_{iN}} ,
\end{align}
where $\mu_B$ is the Bohr magneton, $\mu_0$ is the vacuum permeability, $\gamma_N$ is the gyromagnetic ratio of nucleus $N$, $\vec s_i$ is the spin operator of electron $i$, $e$ is the electronic charge, $m_e$ is the electronic mass, $\vec l_{iN}$ is the orbital angular momentum of electron $i$ with respect to nucleus $N$, and $\vec r_{iN}$ is the vector pointing from nucleus $N$ to electron $i$.

At the level of Hartree-Fock response theory, also known as RPA 
or 
TD-HF,\cite{roweRMP1968} and at the TD-DFT level, 
\cite{Salek2002,Casida2009} the static (i.e., $\omega=0$) linear response function takes the form
\begin{align}
    \langle\langle \hat{P};\hat{O}\rangle\rangle_{\omega=0} = 
    \begin{pmatrix}     
       \mathbf{P}^T(\hat{P}) & -\mathbf{P}^T(\hat{P}) \\
   \end{pmatrix}
     \begin{pmatrix}
      \mathbf{A}  & \mathbf{B} \\
      -\mathbf{B} & -\mathbf{A} \\
     \end{pmatrix}
    ^{-1}
    \begin{pmatrix}
        \mathbf{P}(\hat{O}) \\
        \mathbf{P}(\hat{O}) \\
    \end{pmatrix} .
\end{align}
The dimensions of the $\mathbf{A}$ and $\mathbf{B}$ blocks of the 
electronic 
Hessian matrix 
\begin{align}
\label{eq:Hessian}
\mathbf{E} = \begin{pmatrix}
      \mathbf{A}  & \mathbf{B} \\
      -\mathbf{B} & -\mathbf{A} \\
     \end{pmatrix}
\end{align}
are $N \times N$, with $N$ being the number of single excitations. 
The elements of the $\mathbf{A}$ and $\mathbf{B}$ matrices consist of differences in molecular orbital energies and 
two-electron integrals.
The vectors $\mathbf{P}$ are called the property gradient vectors, whose elements can be evaluated as
\begin{align}
\label{eq:prop_op}
    {P}(\hat{P})_{ai} = \mel{i}{\hat{p}}{a}~,
\end{align}
where $i$ and $a$ denote occupied and virtual spatial orbitals, respectively, and $\hat{p}$ is the one-electron version of $\hat{P}$. 
The inverse of the 
electronic 
Hessian 
matrix is trivially calculated when one knows its eigenvalues and eigenvectors, i.e. after solving the RPA or TD-DFT eigenvalue problem
 \begin{align} 
 \label{eq:Xe-Xd-E-stand}
   \begin{pmatrix}
   \mathbf{A} & \mathbf{B} \\
   -\mathbf{B} & -\mathbf{A} \\
   \end{pmatrix}
   \begin{pmatrix}
       {}^{e}{\mathbf{X}}_n \\
       {}^{d}{\mathbf{X}}_n \\
   \end{pmatrix}
   = \omega_n
   \begin{pmatrix}
       {}^{e}{\mathbf{X}}_n \\
       {}^{d}{\mathbf{X}}_n \\
   \end{pmatrix} ,
 \end{align}
 
\noindent where $^e\mathbf{X}_n$ and $^d\mathbf{X}_n$ are the excitation and de-excitation parts of the eigenvector for the given eigenvalue $\omega_n$, which corresponds to an excitation energy $\Delta E_{0\to n}$ of the system in the RPA/TD-DFT approximation. The right and left eigenvectors of the RPA eigenvalue problem, as written in Eq. \ref{eq:Xe-Xd-E-stand}, are not the same, but the left eigenvectors can be easily generated from the right ones as $\begin{pmatrix} {}^e{\mathbf{X}}_n~~ -{}^d{\mathbf{X}}_n
\end{pmatrix}$.

With these eigenvectors and eigenvalues (excitation energies), one can then rewrite the linear response function expressions for the FC, SD and PSO contributions to the coupling constants as SOS expressions. At the RPA/TD-DFT level the excited states consist of only single-excited states and one can therefore split up the contributions to the isotropic coupling constant into contributions from two pairs of occupied,  $i$ and $j$, and virtual orbitals, $a$ and $b$,
\begin{align}
    J^\textrm{FC/SD/PSO}(N,M) = \sum_{ai, bj} J^\textrm{FC/SD/PSO}_{ai,bj}(N,M)~,
\end{align}
which are defined as
\begin{widetext}
\begin{eqnarray}\label{eq:fc_sd}
\!\!\!\!\!\!\!\!\!\!\!\!\!\!\!\!
J^\textrm{FC/SD}_{ai,bj}(N,M) = - \frac{2}{3}\frac{\gamma_N\gamma_M}{h}
\times \displaystyle\sum_{\alpha=x,y,z}  \displaystyle\sum_n \frac{{P}(\hat{O}^\textrm{FC/SD}_{N\alpha})_{ai}({}^e{X}_{n,ai}-{}^d{X}_{n,ai})({}^e{{X}}_{n,bj}-{}^d{{X}}_{n,bj}){P}(\hat{O}^\textrm{FC/SD}_{M\alpha})_{bj}}{\Delta E_{0\to n}}
\end{eqnarray}
for the FC and SD terms and

\begin{eqnarray}\label{eq:pso}
\!\!\!\!\!\!\!\!\!\!\!\!\!\!\!\!
J^\textrm{PSO}_{ai,bj}(N,M) = \frac{2}{3}\frac{\gamma_N\gamma_M}{h} \times \displaystyle
\sum_{\alpha=x,y,z}\displaystyle\sum_n \frac{{P}(\hat{O}^\textrm{PSO}_{N\alpha})_{ai}(^e{X}_{n,ai}+^d{X}_{n,ai})(^e{{X}}_{n,bj}+^d{{X}}_{n,bj}){P}(\hat{O}^\textrm{PSO}_{M\alpha})_{bj}}{\Delta E_{0\to n}}
\end{eqnarray}
for the PSO term.\cite{sychrovskyJCP2000, spas153} For a closed shell molecule, the FC and SD terms consist of a sum over triplet excited states, while the PSO term is a sum over singlet excited states.\cite{spasB8}
\end{widetext}

\subsection{RPA excitation energies from a Lanczos algorithm}
A more detailed derivation of the equations used in the implementation of the Lanczos algorithm can be found in the paper by Zamok et al.\cite{ZamokJCP2021}
For this section, the focus will be on obtaining the RPA excitation energies with the Lanczos algorithm. 
As anticipated, the 
electronic Hessian matrix in RPA/TD-DFT, as well as e.g. in SOPPA,\cite{spas188}  exhibits a paired structure. This implies that, if $\begin{pmatrix} ^e{\mathbf{X}}_n \ \ {}^d{\mathbf{X}}_n\end{pmatrix}^T$ is an eigenvector for excitation energy $\Delta E_{0\to n}$, then $\begin{pmatrix} ^d{\mathbf{X}}_n \ \ {}^e{\mathbf{X}}_n\end{pmatrix}^T$ will also be an eigenvector of the 
electronic 
Hessian but with an excitation energy  $-\Delta E_{0\to n}$.\cite{OlsenJCoP1988}


To solve the eigenvalue problem, the molecular Hessian
matrix has to be diagonalized. 
The Lanczos algorithm,\cite{lanczosJNBS1950,ZamokJCP2021} 
converges the eigenvalues starting from both ends of the spectrum. 
At the same time one gets, with every iteration, an approximation to another eigenvalue, while at the same time improving the accuracy of the previous eigenvalues. 
This is achieved by iteratively building a tridiagonal representation of the Hessian matrix, $\mathbf{T}^{(N)}$, 
which is afterwards diagonalized. 
The eigenvalues of $\mathbf{T}^{(N)}$ are the same as the eigenvalues of the original matrix $\mathbf{E}$, while the eigenvalues of an intermediate $\mathbf{T}^{(k)}$ matrix, where $k<N$, are approximations to some of the eigenvalues of $\mathbf{E}$. 

For an asymmetric $\mathbf{E}$ matrix, $\mathbf{T}^{(N)}$ is generated in the following manner
\begin{equation}
\label{eq:block-T-RPA}
\tilde{\mathbf{Q}}^{(N)T}\mathbf{E}\mathbf{Q}^{(N)}=\mathbf{T}^{(N)} =
      \begin{pmatrix}
          \mathbf{M}_1 & \tilde{\mathbf{C}}_1^T  &        &  \cdots       &  0           \\
       \mathbf{C}_1  & \mathbf{M}_2 & \ddots &               &  \vdots      \\
                & \ddots   & \ddots &  \ddots       &              \\
       \vdots   &          & \ddots &  \ddots       &  \tilde{\mathbf{C}}_{N-1}^T \\
          0        & \cdots   &        &  \mathbf{C}_{N-1}  &  \mathbf{M}_{N}    \\
      \end{pmatrix} ,
\end{equation}
where $N$ is the dimension of the RPA $\mathbf{A}$ and $\mathbf{B}$ matrices, and $\mathbf{Q}^{(N)}$ is the matrix of the right Lanczos block vectors,
%
\begin{equation}
     \mathbf{Q}^{(N)} = \left( \mathbf{Q}_1 | ... | \mathbf{Q}_{N}  \right)    ;
\end{equation}
similarly, $\mathbf{\tilde{Q}^{(N)}}$ is the matrix of the left Lanczos block vectors:
\begin{equation}
     \tilde{\mathbf{Q}}^{(N)} = 
     \left( \tilde{\mathbf{Q}}_1 | ...| \tilde{\mathbf{Q}}_{N} \right) .
\end{equation}
In our implementation, the matrices of the left and right Lanczos block vectors, $\tilde{\mathbf{Q}}_i$ and $\mathbf{Q}_i$, are constructed to have the same symmetry as the left and right RPA (excitation and de-excitation) eigenvectors, i.e., they are defined as:
\begin{align}
      \label{eq:blockL-def}
       \mathbf{Q}_i
        \equiv
      \begin{pmatrix}
        \mathbf{X}_i & \mathbf{Y}_i \\
        \mathbf{Y}_i & \mathbf{X}_i
      \end{pmatrix}
%
\text{  and  }      
%
        \tilde{\mathbf{Q}}_i
        \equiv
      \begin{pmatrix}
        \mathbf{X}_i & -\mathbf{Y}_i \\
        -\mathbf{Y}_i & \mathbf{X}_i
      \end{pmatrix}
\end{align}
in order to preserve the RPA paired structure also in the Lanczos basis.

As the dimensions of the $\mathbf{X}_i$ and $\mathbf{Y}_i$ are
    \begin{align}
      \mathbf{X}_i, \mathbf{Y}_i \in \mathbb{R}^{N \times 1}  ,
    \end{align}
the dimensions of the Lanczos block vectors are
\begin{equation}
     \mathbf{Q}_i, \tilde{\mathbf{Q}}_i \in \mathbb{R}^{2N \times 2}
\end{equation}
and
\begin{equation}
     \mathbf{Q}^{(N)}, \tilde{\mathbf{Q}}^{(N)}, \mathbf{E} \in \mathbb{R}^{2N \times 2N} .     
\end{equation}
Consequently, the dimensions of submatrices in the full block-tridiagonal matrix $\mathbf{T}^{(N)}$ are
 \begin{equation}
     \mathbf{M}_i, \mathbf{C}_i, \tilde{\mathbf{C}}_i \in \mathbb{R}^{2 \times 2} .
 \end{equation}

The matrices $\mathbf{M}_{k}$, where $k$ denotes the iteration number, are constructed by transforming the $\mathbf{E}$ matrix with a left and right Lanczos block vector. As it will be shown later, the result then exhibits a paired structure of excitation $e_k$ and de-excitation elements $d_k$,
\begin{align}
        \label{eq:m_mat}
        \mathbf{M}_{k} 
         = \tilde{\mathbf{Q}}_k^T \mathbf{E}\mathbf{Q}_k 
         = \tilde{\mathbf{Q}}_k^T\mathbf{R}_{k} 
        \equiv
          \begin{pmatrix}
           e_k & d_k \\
          -d_k & -e_k     
          \end{pmatrix} .
\end{align}
The two Lanczos block vectors must be bi-orthonormal, which means that $\tilde{\mathbf{Q}}_1^T\mathbf{Q}_1=\mathbf{I_2}$, where $\mathbf{I_2}$ is the $2\times2$ identity matrix. 
The normalization of the Lanczos block vectors is thus a consequence of the normal RPA normalization $\mathbf{X}_k^T\mathbf{X}_k-\mathbf{Y}_k^T\mathbf{Y}_k=1$. 
When stopping after $k$ iterations, a block tridiagonal\footnote{In reality, in the algorithm presented here the matrix is hexa-diagonal.} matrix $\mathbf{T}^{(k)}$ is obtained. 
This matrix thus has an eigenspectrum which will be an approximation to the one of $\mathbf{E}$. 

Because of the bi-orthonormality requirement, the recursion for creating the new right Lanczos block at the $k^{th}$ iteration is given as
\begin{align}
          \label{eq:impl-block-recursion-right}
       \mathbf{R}_{k} &= \mathbf{E}\mathbf{Q}_k =
          \begin{pmatrix}
        \mathbf{A} \mathbf{X}_k + \mathbf{B}\mathbf{Y}_k  && \mathbf{A} \mathbf{Y}_k + \mathbf{B}\mathbf{X}_k \\
        -\mathbf{B}\mathbf{X}_k - \mathbf{A}\mathbf{Y}_k  && -\mathbf{B}\mathbf{Y}_k - \mathbf{A} \mathbf{X}_k
      \end{pmatrix} \nonumber \\
        &\equiv
          \begin{pmatrix}
              \mathbf{X}_{k+1}'  & -\mathbf{Y}_{k+1}'  \\
              \mathbf{Y}_{k+1}'  & -\mathbf{X}_{k+1}'
          \end{pmatrix} .
\end{align}
From the new right and left Lanczos blocks, the two previous blocks are projected out as in the following:
\begin{align}          \label{eq:QR_k}
          &\mathbf{Q}_{k+1} \mathbf{C}_k
          = 
          \mathbf{R}_{k}' 
          = 
          \mathbf{R}_k 
          - \mathbf{Q}_{k} \mathbf{M}_{k} 
          - \mathbf{Q}_{k-1} \tilde{\mathbf{C}}^T_{k-1}
          \\
          \label{eq:QS_k}
          &\tilde{\mathbf{Q}}_{k+1} \tilde{\mathbf{C}}_k
           = 
           \mathbf{S}_{k}' 
           = 
           \mathbf{S}_{k} 
           - \tilde{\mathbf{Q}}_{k}\mathbf{M}_{k}^T 
           - \tilde{\mathbf{Q}}_{k-1}\mathbf{C}_{k-1}^T
\end{align}
The super-diagonal coefficients blocks, $\mathbf{C}_{k-1}^T$ are defined as
\begin{widetext}
\begin{align}
       \label{eq:C-tilde-block}
         \tilde{\mathbf{C}}^T_{k-1} 
        = 
        \tilde{\mathbf{Q}}^T_{k-1} \mathbf{E} \mathbf{Q}_k 
        = 
        \tilde{\mathbf{Q}}^T_{k-1} \mathbf{R}_k  
        =
          \begin{pmatrix}
            \mathbf{X}_{k-1}^T\mathbf{X}_{k+1}' -  \mathbf{Y}_{k-1}^T\mathbf{Y}_{k+1}'  
             &  - \mathbf{X}_{k-1}^T\mathbf{Y}_{k+1}' + \mathbf{Y}_{k-1}^T\mathbf{X}_{k+1}'     
             \\ 
             -\mathbf{Y}_{k-1}^T\mathbf{X}_{k+1}' + \mathbf{X}_{k-1}^T\mathbf{Y}_{k+1}'  
             & \mathbf{Y}_{k-1}^T\mathbf{Y}_{k+1}' - \mathbf{X}_{k-1}^T\mathbf{X}_{k+1}'
           \end{pmatrix} 
          \equiv
          \begin{pmatrix}
            a_{k-1} & b_{k-1} \\
           -b_{k-1} & -a_{k-1}
          \end{pmatrix}
          \end{align}
\end{widetext}
and the sub-diagonal coefficient blocks, $\mathbf{C}_{k-1}$, can in turn be shown to be equal to the super-diagonal coefficient blocks
\begin{align}           
\label{eq:C-block}
         & \mathbf{C}_{k-1} 
         = 
         (\mathbf{Q}_{k-1}^T \mathbf{E}^T \tilde{\mathbf{Q}}_{k})^T
         = 
         (\mathbf{Q}_{k-1}^T \mathbf{S}_k)^T
         =
         \tilde{\mathbf{C}}^T_{k-1} .
\end{align}

The case of the left Lanczos block, $\mathbf{S}_{k}$, is very similar, and the recursion for creating it is given as
\begin{align}  \label{eq:impl-block-recursion-left}
          \mathbf{S}_{k} &= \mathbf{E}^T\tilde{\mathbf{Q}}_k =
          \begin{pmatrix}
        \mathbf{A} \mathbf{X}_k + \mathbf{B}\mathbf{Y}_k  && -\mathbf{A} \mathbf{Y}_k - \mathbf{B}\mathbf{X}_k \\
        \mathbf{B}\mathbf{X}_k + \mathbf{A}\mathbf{Y}_k  && -\mathbf{B}\mathbf{Y}_k - \mathbf{A} \mathbf{X}_k
      \end{pmatrix} \nonumber \\
        &\equiv
          \begin{pmatrix}
              \mathbf{X}_{k+1}'   &  \mathbf{Y}_{k+1}'  \\
              -\mathbf{Y}_{k+1}'  & -\mathbf{X}_{k+1}'
          \end{pmatrix} .
\end{align}

The symmetry between the right and left Lanczos vectors $\mathbf{R}_{k}$ and $\mathbf{S}_{k}$ is left unchanged and therefore one only needs to do the recursion for one of the new unnormalized Lanczos blocks $\mathbf{R}'_{k}$ and $\mathbf{S}'_{k}$, because they only contain one linearly dependent Lanczos vector. Furthermore, it is sufficient to only do the recursion for the first vector in the new unnormalized right Lanczos block $\mathbf{Q}_{k+1}$, since it is then already known how the second will look like. 
%
\begin{align}
    \label{eq:pseudocode-recursion}
    (\mathbf{R}_k')_1 
    &\equiv
    \begin{pmatrix}
        \mathbf{X}_{k+1}''
        \\
        \mathbf{Y}_{k-1}''
    \end{pmatrix} \\
    &=
    \begin{pmatrix}
            \mathbf{X}_{k+1}'
            -e_k\mathbf{X}_{k}       +d_k\mathbf{Y}_k 
            -a_{k-1}\mathbf{X}_{k-1} +b_{k-1}\mathbf{Y}_{k-1}
            \\
            \mathbf{Y}_{k+1}'
            -e_k\mathbf{Y}_{k}       +d_k\mathbf{X}_k  
            -a_{k-1}\mathbf{Y}_{k-1} +b_{k-1}\mathbf{X}_{k-1}
    \end{pmatrix} .
\end{align}

The product of $\mathbf{S}'^T_{k}$ and $\mathbf{R}'_{k}$, given in Eq. \eqref{eq:B^TB}, is needed to obtain the inverse of the $\tilde{\mathbf{C}}_k$ and $\mathbf{C}_k$. These two are needed in order to normalize the new Lanczos vector.
\begin{align}
            \label{eq:B^TB}
            \mathbf{S}_k^{'T} \mathbf{R}_{k}' = 
            \tilde{\mathbf{C}}_{k}^T\mathbf{C}_{k} =
            \begin{pmatrix}
                a_{k}^2 - b_{k}^2 & 0   \\
                 0                    & a_{k}^2 - b_{k}^2
       \end{pmatrix}
            =
            \begin{pmatrix}
                \mathcal{N}_{k+1} & 0   \\
                 0 & \mathcal{N}_{k+1}
       \end{pmatrix}
\end{align}
$\mathcal{N}_{k+1}$ is the normalization constant for the $(k+1)^{th}$ Lanczos block squared. The values of $a_k$ and $b_k$ are not uniquely defined and therefore, to simplify things, one of them is set to zero. From looking at the product of $\tilde{\mathbf{C}^T_k}\mathbf{C}^T_k$, it can be seen that, if $a_k$ is set to zero, then $\mathcal{N}_{k+1}$ will be negative, and if $b_k$ is set to zero then it will be positive. This is done in the following manner:
 \begin{align}   
 \label{eq:ab-scheme}
   &\mathcal{N}_{k+1} > 0: \quad a_{k} = \sqrt{\mathcal{N}_{k+1}} \ , \ \ \quad b_k = 0 
   \\      
   &\mathcal{N}_{k+1} < 0: \quad b_{k} = \sqrt{-\mathcal{N}_{k+1}} \ , \quad a_k = 0    \nonumber
 \end{align}

To preserve the bi-orthogonality that can be lost due to round-off errors, each new Lanczos vector is explicitly re-orthogonalised: 
\begin{align}
       \label{eq:bi-orthog}
           \begin{pmatrix}
             \mathbf{X}_{k+1}'' \\
             \mathbf{Y}_{k+1}''
            \end{pmatrix}
            =
           \begin{pmatrix}
             \mathbf{X}_{k+1}' \\
             \mathbf{Y}_{k+1}'
            \end{pmatrix}
          -
           \mathbf{Q}^{(k)}\tilde{\mathbf{Q}}^{(k)T}
            \begin{pmatrix}
             \mathbf{X}_{k+1}' \\
             \mathbf{Y}_{k+1}'
            \end{pmatrix}
    \end{align}

The same symmetry as in the RPA Hessian $\mathbf{E}$ can also be obtained for the $\bar{\mathbf{T}}^{(k)}$ matrix by reordering it after $k$ iterations:
  \begin{align}   \label{eq:Tbar-mat}
   &\begin{pmatrix}
       \mathbf{U}^{(k)T} & -\mathbf{V}^{(k)T} \\
       -\mathbf{V}^{(k)T} & \mathbf{U}^{(k)T}
   \end{pmatrix}
   \begin{pmatrix}
    \mathbf{A} & \mathbf{B} \\
    -\mathbf{B} & -\mathbf{A}
   \end{pmatrix}
   \begin{pmatrix}
       \mathbf{U}^{(k)} & \mathbf{V}^{(k)} \\
       \mathbf{V}^{(k)} & \mathbf{U}^{(k)}
   \end{pmatrix} \\
   &=
   \begin{pmatrix}
    \mathbf{A}^{'(k)} & \mathbf{B}^{'(k)} \\
    -\mathbf{B}^{'(k)} & -\mathbf{A}^{'(k)}
   \end{pmatrix}
   =
   \bar{\mathbf{T}}^{(k)} ,
  \end{align}
where the $i^{th}$ column in $\mathbf{U}^{(k)}$ consists of the upper part of the first vector of $\mathbf{Q}_i$, i.e.
\begin{align}
     &\mathbf{U}^{(k)} = [ \mathbf{X}_1,..,\mathbf{X}_k ] ,
\end{align}
and the $i^{th}$ column in $\mathbf{V}^{(k)}$ is the lower part of the first vector of $\mathbf{Q}_i$, i.e.
\begin{align}
        \mathbf{V}^{(k)} = [ \mathbf{Y}_1,..,\mathbf{Y}_k ] .
\end{align}
%
%
%
%
After $k$ iterations, the dimensions of the $\mathbf{U}^{(k)}$ and $\mathbf{V}^{(k)}$ matrices are thus given as:
    \begin{align}
     \mathbf{U}^{(k)}, \mathbf{V}^{(k)} \in \mathbb{R}^{N \times k}
    \end{align}
and correspondingly for the $\mathbf{A}'^{(k)}$ and $\mathbf{B}'^{(k)}$ matrices:
    \begin{align}
     \mathbf{A}^{\prime(k)}, \mathbf{B}^{\prime(k)} \in \mathbb{R}^{k \times k} .
     \end{align}

Getting the RPA eigenvalues then becomes a question of solving the eigenvalue problem for the reordered $\bar{\mathbf{T}}^{(k)}$ matrix:
 \begin{align} \label{eq:T_eig}
   \begin{pmatrix}
   \mathbf{A}^{\prime(k)} & \mathbf{B}^{\prime(k)}  \\
   -\mathbf{B}^{\prime(k)}  & -\mathbf{A}^{\prime(k)}  \\
   \end{pmatrix}
   \begin{pmatrix}
       {}^{e}{\bm{\Xi}}_n \\
       {}^{d}{\bm{\Xi}}_n \\
   \end{pmatrix}
   =   \omega_n
   \begin{pmatrix}
       {}^{e}{\bm{\Xi}}_n \\
       {}^{d}{\bm{\Xi}}_n \\
   \end{pmatrix} ,
 \end{align}
where $^e\bm{\Xi}_n$ and $^d\bm{\Xi}_n$ are the RPA eigenvectors in the basis of the Lanczos vectors and have thus dimensions of $^e\bm{\Xi}_n$ and $^d\bm{\Xi}_n$:
    \begin{align}
     ^e\bm{\Xi}_n, ^d\bm{\Xi}_n \in \mathbb{R}^{k} .
     \end{align}

In the previous study by Zamok et al.\cite{ZamokJCP2021} the aim was to calculate mean excitation energies, and for those calculations the eigenvectors of $\bar{\mathbf{T}}^{(k)}$ were sufficient. 
However, for the analysis of the SSCCs in terms of contributions from individual pairs of occupied and virtual orbitals, i.e. Eqs. (\ref{eq:fc_sd}) and (\ref{eq:pso}), this is not sufficient and we have to generate therefore eigenvectors in the full space in the following manner:
\begin{align}
    \begin{pmatrix}
        ^e\mathbf{X}_{n}        \\
        ^d\mathbf{X}_{n}
    \end{pmatrix}
    =
    \begin{pmatrix}
            \displaystyle\sum_k \left [ \mathbf{X}_{k}  (^e\bm{\Xi}_{n})_k + \mathbf{Y}_k  (^d\bm{\Xi}_{n})_k \right ]
            \\
            \displaystyle\sum_k \left [ \mathbf{X}_{k}  (^d\bm{\Xi}_{n})_k + \mathbf{Y}_k (^{e}\bm{\Xi}_{n})_k \right ]
    \end{pmatrix} ,
\end{align}
where $\mathbf{X}_{k}$ and  $\mathbf{Y}_k$ are the excitation and de-excitation parts of the $k^{th}$ Lanczos vector, and $^e\mathbf{X}_{n}$ and $^d\mathbf{X}_{n}$ are the excitation and de-excitation part of the $n^{th}$ full space eigenvector. This can be achieved as shown in Algorithm \ref{alg:eigv_trans}.
%
\begin{algorithm}[H]
\DontPrintSemicolon
    \KwData{$[(^e\bm{\Xi}_{n})_1]\dots[(^e\bm{\Xi}_{n})_{k_{max}}], [(^d\bm{\Xi}_{n})_1]\dots[(^d\bm{\Xi}_{n})_{k_{max}}] \in \mathbb{R}^{N}$, 
    $[X_1]\dots[X_{k_{max}}], [Y_1]\dots[Y_{k_{max}}]  \in \mathbb{R}^{N}$,  $k_{max}$ Lanczos chain size.}
    \KwResult{$[^e\mathbf{X}_{1}]\dots[^e\mathbf{X}_{N}],[^d\mathbf{X}_{1}]\dots[^d\mathbf{X}_{N}] \in \mathbb{R}^{N}$}
\Begin{
    $^e\mathbf{X}_{1}=0$ \;
    $^d\mathbf{X}_{1}=0$ \;
    $k = 1$ \;
    \While{$k < k_{max}$}{
        ${}^e\mathbf{X}_{k}=X_k(^e\bm{\Xi}_{n})_k+Y_k(^d\bm{\Xi}_{n})_k$ \;
        ${}^d\mathbf{X}_{k}=X_k(^d\bm{\Xi}_{n})_k+Y_k(^e\bm{\Xi}_{n})_k$ \;
        \;
        $\mathcal{N}_{k}={}^e\mathbf{X}_{k}{}^e\mathbf{X}_{k}-{}^d\mathbf{X}_{k}{}^d\mathbf{X}_{k}$ \;
        $\mathcal{N}_k = \sqrt{\frac{0.5}{\mathcal{N}_k}}$ \;
        \;
        $^e\mathbf{X}_{k}=-X_k(^e\bm{\Xi}_{n})_k\mathcal{N}_{k}$ \;
        $^d\mathbf{X}_{k}=-X_k(^d\bm{\Xi}_{n})_k\mathcal{N}_{k}$ \;
        $k=k+1$ \;
}
}
\caption{The transformation of the eigenvectors to full space} \label{alg:eigv_trans}
\end{algorithm}

A final point to be discussed is the choice of start vectors for the Lanczos algorithm 
(such that we conserve the paired structure and obtain excited states that are relevant for the desired properties).
In the previous study by Zamok et al.,\cite{ZamokJCP2021} the 
property gradient vectors $\mathbf{P}$, Eq. \eqref{eq:prop_op}, of the electric dipole operator were employed, as the desired property anyway involved electric dipole transition moments. 
Furthermore, it was previously shown that one can then calculate the corresponding response property directly in the Lanczos base.\cite{hansenJCP2010,corianiJCTC2012}
In the present case of SSCCs, 
the situation is more complicated as we deal with excitation energies to both singlet and triplet states and, for the latter case, it ought to be integrals of a perturbation operator involving spin as, e.g., the Fermi-contact or spin-dipole operators, Eqs. \eqref{eq:o_fc} and \eqref{eq:o_sd}. 
Thus, several options seem possible and will be specifically discussed in Section~\ref{Results}. 

\section{Computational details}
\label{CompDetails}
The new Lanczos algorithm, implemented in a local version of the DALTON program package,\cite{daltonpaper} was used to calculate the Fermi-contact contribution to the SSCCs for the 17 molecules BH$_3$, C$_2$H$_2$, C$_2$H$_4$, C$_2$H$_6$, CH$_4$, CO, 
H$_2$CO, H$_2$O, H$_2$S, HCN, HCl, HF, N$_2$ N$_2$O, NH$_3$, PH$_3$ and SiH$_4$.
Some were selected, because they were the test molecules in another study,\cite{spas153} where  it was analogously investigated (at the DFT level) how many excited states obtained with the Davidson algorithm are necessary to converge the SSCCs.
Including these molecules in our test set allows us to compare directly with this previous study in the following.
The compounds in question are CH$_4$, NH$_3$, H$_2$O, SiH$_4$, PH$_3$, H$_2$S, C$_2$H$_2$, C$_2$H$_4$ and C$_2$H$_6$. 
To get an even wider picture of how the algorithm performs, the molecules BH$_3$, CO, 
H$_2$CO, HCN, N$_2$, N$_2$O, HCl and HF were also included in the test set. 
They were used in the recent study on mean excitation energies with the Lanczos algorithm.\cite{ZamokJCP2021}

The geometries have been previously optimized using the correlation-consistent basis set \mbox{cc-pVTZ}\cite{cc-pCVnZ-Dunning-H-BtoNe,cc-pCVnZ-Dunning-AltoAr} at the CCSD(T) level of theory.\cite{Sauer-I0-molecular-ions,Sauer-I0-states,Sauer-I0-2019,Sauer-I0-geom}
The SSCC calculations were carried out with the triple zeta polarization-consistent basis set pcJ-2,\cite{JensenTCA2010} which is specifically optimized for the calculation of SSCCs. 
For each molecule, several calculations were done, whose purpose was to investigate how fast the SSCCs converge with the number of approximate states included in the sum-over-states expressions, Eqs. \eqref{eq:fc_sd} and \eqref{eq:pso}, i.e., with the Lanczos chain length. 
As a reference we employed the value of the SSCCs calculated as a linear response function, i.e. from Eq. (\ref{eq:sscc_lr}). All calculations were carried out at the RPA level. This implies that one should not expect good agreement with experimental values. In the context of the present work, however, this is irrelevant, as the convergence pattern can be expected to be the same at the DFT level of theory.

When the maximum number of excitation energies was less than 1000, we started the calculations using a Lanczos chain length of 20, and then increased the chain length by steps of 10, until the maximum was reached. 
When the maximum number of excitation energies was more than 1000, the calculations started with a chain length of 50 and then increased stepwise by 50, until the maximum was reached. 
For the C$_2$H$_6$ molecule we started at a chain length of 50 with an increase of 50 until a chain length of 1600 was reached. From there, each new calculation increased the chain length by 100 until the maximum was reached. 
The values of the coupling constants as a function of the chain length were then plotted using a python script.

\section{Results and Discussion}
\label{Results}
The purpose of this study is to investigate how fast the Fermi-contact contribution to the SSCCs calculated as a sum-over-states converge with the Lanczos chain length, i.e. the number of Lanczos iterations.
To do this, a number of test molecules was chosen. 
The couplings present in the total set of molecules can be put into four different categories: one-bond couplings, two bond couplings, geminal couplings, and vicinal couplings. 
For most of the couplings, the most dominating term is the Fermi-Contact term. 

\subsection{One-bond couplings}

In the set of molecules examined in this study, two categories of one‑bond couplings are present. The first involves X–H couplings, where X = B, C, N, O, F, Si, P, S, or Cl. The second consists of X–Y couplings, where X and Y = C, N, or O. Both categories cover a broad span of coupling‑constant magnitudes.
For example, C–H couplings alone range from 155 to 439 Hz, while other X-H couplings extend from $-247$ Hz (Si-H) to 659 Hz (F-H). Among the X-Y couplings, C-C values fall between 60 and 424 Hz, whereas the remaining X-Y couplings vary from $-110$ Hz (C-O in H$_2$CO) to 309 Hz (N-O in N$_2$O). Thus, although these coupling types may appear quite different, their numerical ranges overlap substantially. More strictly considering only the absolute values, X–H couplings span from 32 Hz (Cl–H) to 659 Hz (F–H), whereas X–Y couplings range from 4.5 Hz (C–O in CO) to 424 Hz (C–C in C$_2$H$_2$), indicating that the spread in X–Y couplings is comparatively narrower.

The largest term in the majority of one-bond couplings is the FC term. 
We will therefore concentrate only on this contribution. 
However, there are cases where also the other contributions, SD and PSO, have an influence on the total coupling constants, implying that there will be differences between the FC term and the total coupling constant. For the SD contribution, these  
are the N-N couplings in N$_2$O and in N$_2$, the N-O coupling in N$_2$O, the C-O coupling in CO and H$_2$CO, the C-N coupling in HCN, and the C-C coupling in ethene. 
For the PSO term, these are the C-O couplings in CO and H$_2$CO, the H-F and H-Cl couplings in HF and HCl, as well as the H-S and H-O couplings in H$_2$S and H$_2$O.


\begin{figure}[!htb]
    \centering
    \includegraphics[width=0.9\linewidth]{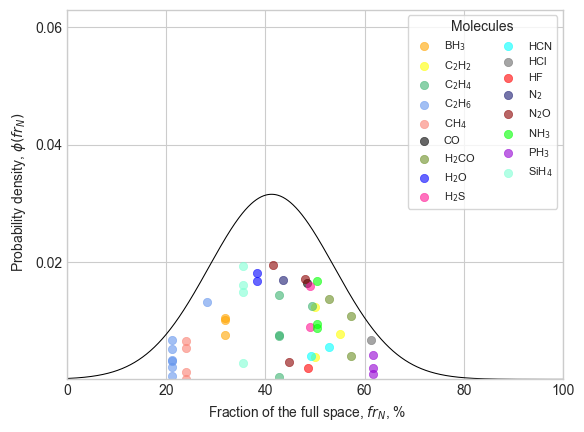}
    \caption{The fraction fr$_N$ of the full space of excitations at which the Fermi-contact terms, $^1J^\textrm{FC}(M,N)$), converge with an allowed deviation of 0.5 Hz for all one-bond couplings.}
    \label{fig:onebond_fc_hz}
\end{figure}
In Figure \ref{fig:onebond_fc_hz}, results for all the one-bond couplings in all molecules of this study are collected. 
The figure shows at which fraction fr$_N$ of the full space of excitations the, Fermi-contact contribution to the couplings is converged within a threshold of 0.5 Hz.
Note that the y-value of the individual points is of no importance, since the points of the different molecules have been spread out in order to make them easier to distinguish. 
The y-axis refers only to the probability density function, which is shown as the solid line. 
The figure shows that the Fermi-contact term for most molecules converges when around a fr$_N$ of 41$\%$ is used in the calculation; for some couplings, it converges already at around 20$\%$ and for others up to a little over 60$\%$ of the full space is needed for convergence of the FC term.

\begin{figure}[h!bt]
\centering
  \includegraphics[width=0.9\linewidth]{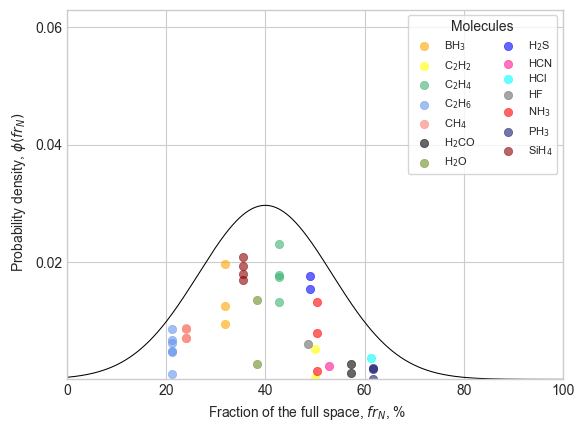}
\caption{The fraction of the full space fr$_N$ at which the Fermi-contact terms converge with an allowed deviation of 0.5 Hz for all X-H one-bond couplings.}
\label{fig:one-bond_X-HandX-Y-1}
\end{figure}

\begin{figure}[h!]
  \centering
  \includegraphics[width=0.9\linewidth]{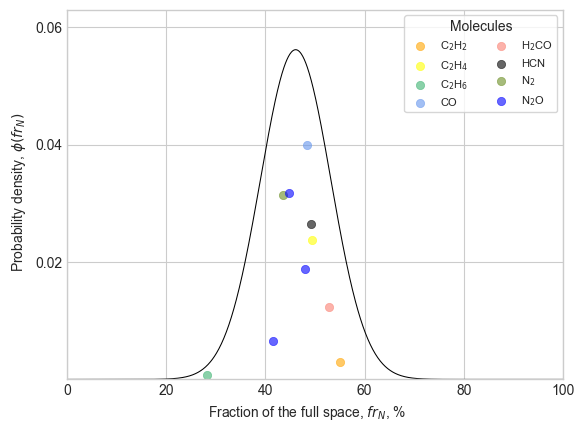}
\caption{The fraction of the full space fr$_N$ at which the Fermi-contact terms converges with an allowed deviation of 0.5 Hz for all X-Y one-bond couplings.}
\label{fig:one-bond_X-HandX-Y-2}
\end{figure}

While Figure \ref{fig:onebond_fc_hz} contains two types of one-bond couplings, X-H and X-Y, in Figures \ref{fig:one-bond_X-HandX-Y-1} and \ref{fig:one-bond_X-HandX-Y-2} we have split the two types into two different plots. 
Figure \ref{fig:one-bond_X-HandX-Y-1} shows all the X-H couplings, and Figure \ref{fig:one-bond_X-HandX-Y-2} all the X-Y (and X-X) couplings. 
The plots show that both types of couplings converge at around 42$\%$ (40$\%$ for X-H and 46$\%$ for X-Y), whereas the X-H couplings are the ones with the bigger spread. 
The coupling constants that need a fr$_N$ of 21.1$\%$ to converge are the six C-H couplings in ethane, 
and the coupling constants that need a fr$_N$ of 62$\%$ to converge are the P-H couplings in PH$_3$.

As anticipated in the theory section, one needs to choose a start vector for the Lanczos algorithm, and a preferred choise is a property gradient vector related to the desired response property. 
In the case of the FC contribution to the SSCCs, this could be the property gradient vector for the FC operator of one of the two coupled nuclei. 
The results shown in Figure \ref{fig:onebond_fc_hz}, as well as in Figures \ref{fig:one-bond_X-HandX-Y-1} and \ref{fig:one-bond_X-HandX-Y-2}, are the ones obtained with the FC property gradient vector, which gave the fastest convergence. 


For the acetylene C-H couplings, both the C1-H1 and the C2-H2 couplings are given in the figure since calculations with both carbon FC property gradients were performed. 
For ethene and ethane, 
only the couplings between C1 and the hydrogen atoms are given in the figures. 
The C-O coupling constants in CO 
and H$_2$CO converged fastest with the carbon FC property gradients as start vectors (compared to those obtained using the oxygen FC property gradients), therefore it is the results of the calculations with the carbon FC property gradients that are shown in the figures. 

The N$_2$O molecule is an especially interesting case. 
When the FC property gradient corresponding to the central nitrogen atom is used, the results never reach the linear response value; the final results deviate by 0.5 Hz for the N-N coupling and 1.0 Hz for the N-O one-bond coupling constant. 
On the other hand, when the FC property gradient of the terminal nitrogen is used for 
the N-N coupling constant, the final deviation is only less than 0.1 Hz and; 
when the oxygen FC property gradient is used to calculate the N-O one-bond coupling constant, the deviation is less than 0.1 Hz.
For none of the X-H one-bond coupling constants, the results were better with the hydrogen FC property gradients. However, the H-Cl coupling deviated even with the chlorine FC property gradient by 0.1 Hz from the linear response result, when the full space was used.

\begin{figure}[h!]
\centering
  \includegraphics[width=1.\linewidth]{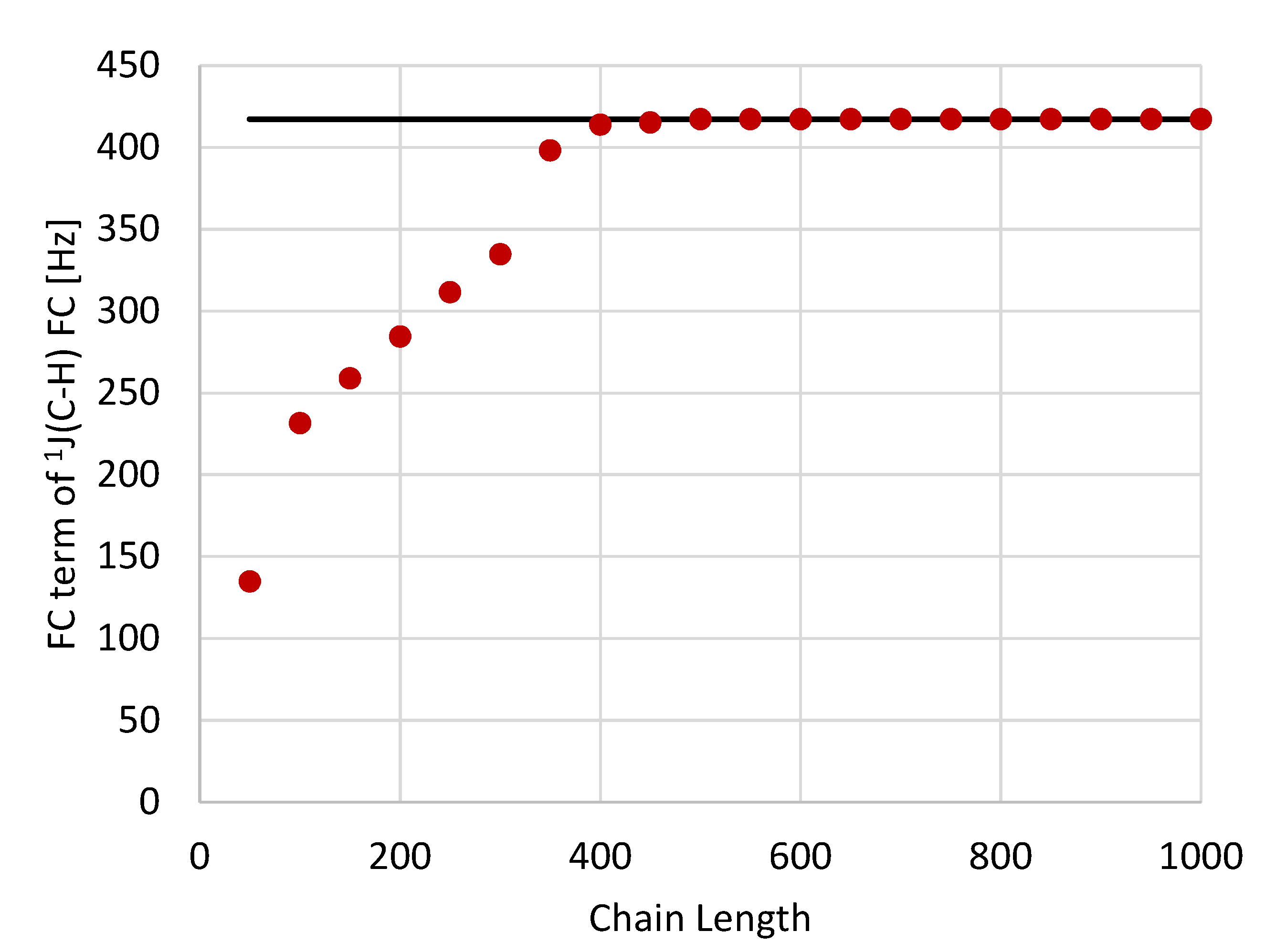}
\caption{C$_2$H$_2$: The Fermi-contact term for the one-bond coupling between C1 and H1, $^1J^{\textrm{FC}}$(C1-H1), as a function of the Lanczos chain length. The calculations were carried out with the property gradient for the FC operator of C1 as start vector.
The reference value of the FC term calculated as linear response function is shown as solid line.}
\label{fig:C2H2_H1-C1}
\end{figure}
\begin{figure}[hbtp!]
\centering
  \includegraphics[width=1.\linewidth]{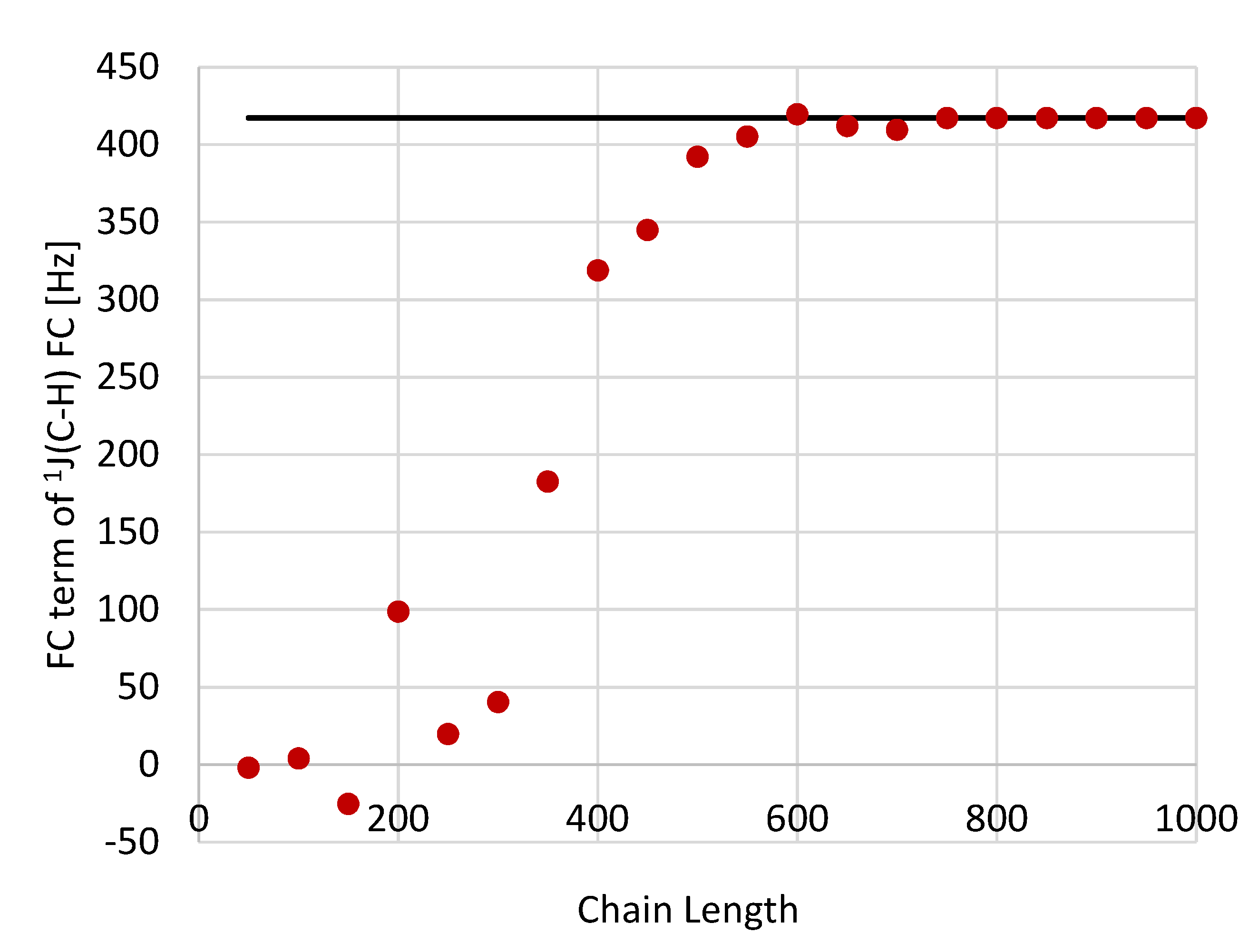}
\caption{C$_2$H$_2$: The Fermi-contact term for the one-bond coupling between C2 and H2, $^1J^{\textrm{FC}}$(C2-H2), as a function of the Lanczos chain length. The calculations were carried out with the property gradient for the FC operator of C1 as start vector.
The value of the FC term calculated as linear response function is shown as solid line.}
\label{fig:C2H2_H2-C2}
\end{figure}

For most of the coupling constants, the FC term follows approximately the same convergence pattern, but not all degenerate couplings converge with the same speed or pattern. 
As previously mentioned, the rate of convergence depends on the choice of start vector. To illustrate this, we consider the case of the two equivalent C-H couplings in acetylene, shown in Figures~\ref{fig:C2H2_H1-C1} and \ref{fig:C2H2_H2-C2}, where we have used the C1 FC property gradient as start vector. 
The end result for both of them is 417 Hz, but they converge at very different rates. 
The C2-H2 coupling, shown in Figure~\ref{fig:C2H2_H2-C2}, oscillates much more initially than the C1-H1 coupling.
This issue can be solved by changing the property gradient used as start vector for the calculation of the triplet excitation energies. 
The FC term becomes better if the FC property gradient corresponding to the coupling is chosen. 
For acetylene, the first calculation in Figure \ref{fig:C2H2_H1-C1} was performed using the FC integral for C1, therefore the couplings between C1 and other atoms converge faster than the couplings with C2. 
On the other hand, when the FC property gradient for C2 is used, all the couplings with C2 converge fast, and the convergence pattern for C2-H2 now looks exactly like C1-H1 did and vice versa.
However, it is not only the coupling over the two C-H bonds in acetylene that converges at different chain lengths. The same is true for other equivalent bonds such as in the molecules C$_2$H$_4$ and C$_2$H$_6$. 
Therefore, using as start vector the corresponding type of property gradients for the different couplings and terms is of great importance.

\begin{figure}[!h]
\centering
  \includegraphics[width=1\linewidth]{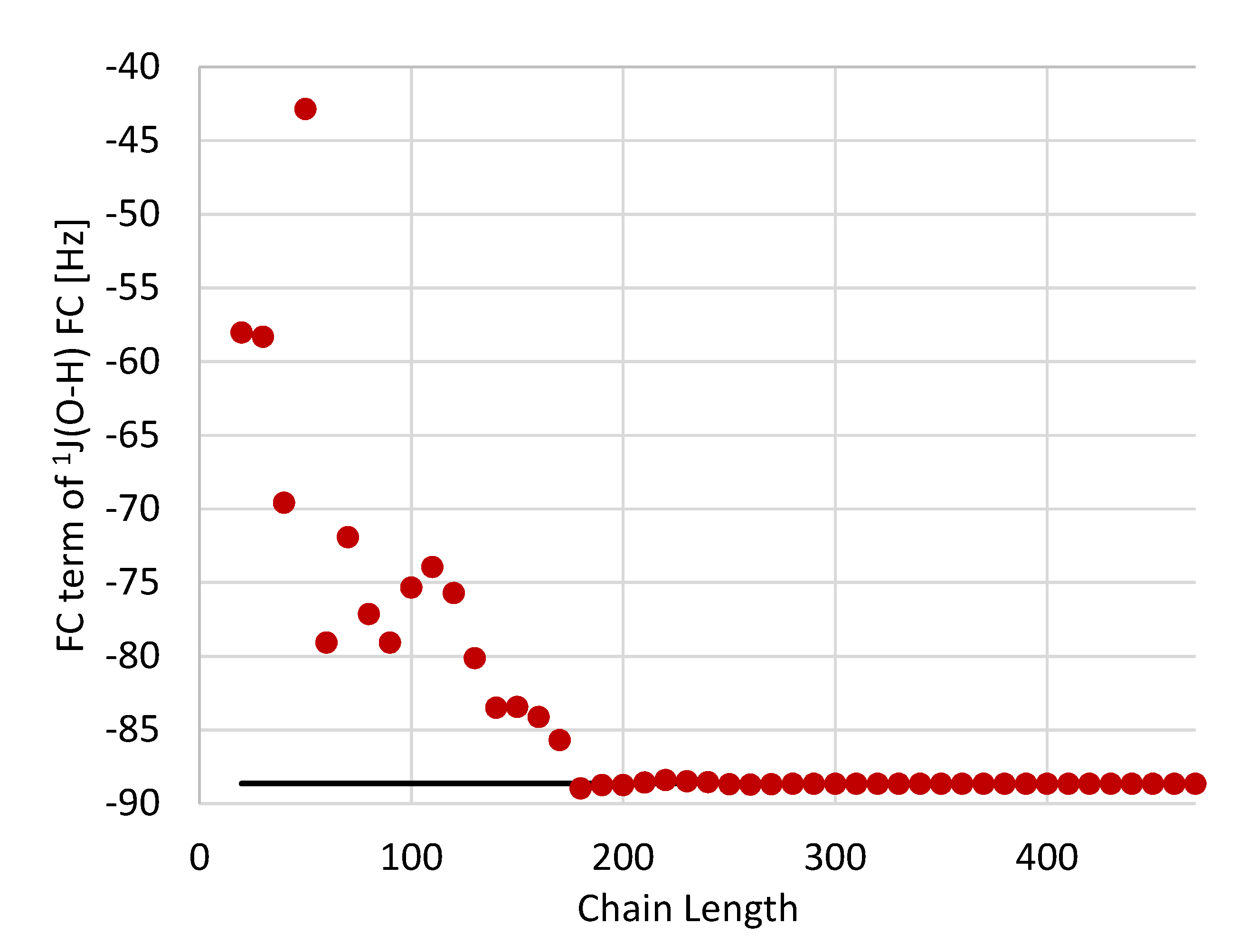}
\caption{H$_2$O: The Fermi-contact term for the one-bond coupling between O and H, $^1J^{\textrm{FC}}$(O-H), as a function of the Lanczos chain length. The calculations were carried out with the property gradient for the FC operator of O as start vector.
The value of the FC term calculated as linear response function is shown as solid line.}
\label{fig:H2O_H1-O1}
\end{figure}

\begin{figure}[!htb]
\centering
  \includegraphics[width=1\linewidth]{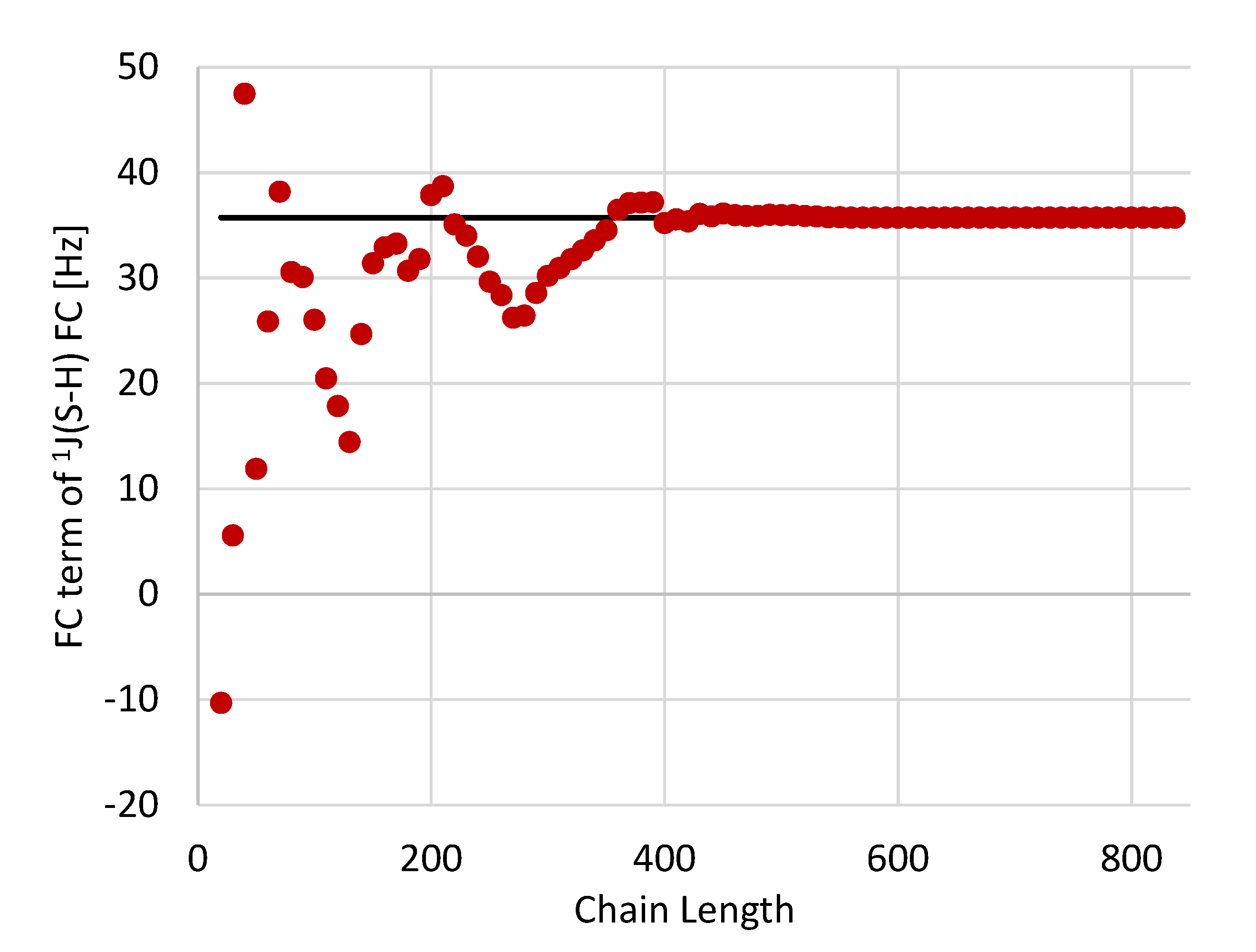}
\caption{H$_2$S:  The Fermi-contact term for the H-S one-bond coupling, $^1J^{\textrm{FC}}$(S-H), as a function of the Lanczos chain length. The calculations were carried out with the property gradient for the FC operator of S as start vector.
The value of the FC term calculated as linear response function is shown as solid line.
}
\label{fig:H2S_H1-S1}
\end{figure}

\begin{figure}[!htb]
\centering
  \includegraphics[width=1\linewidth]{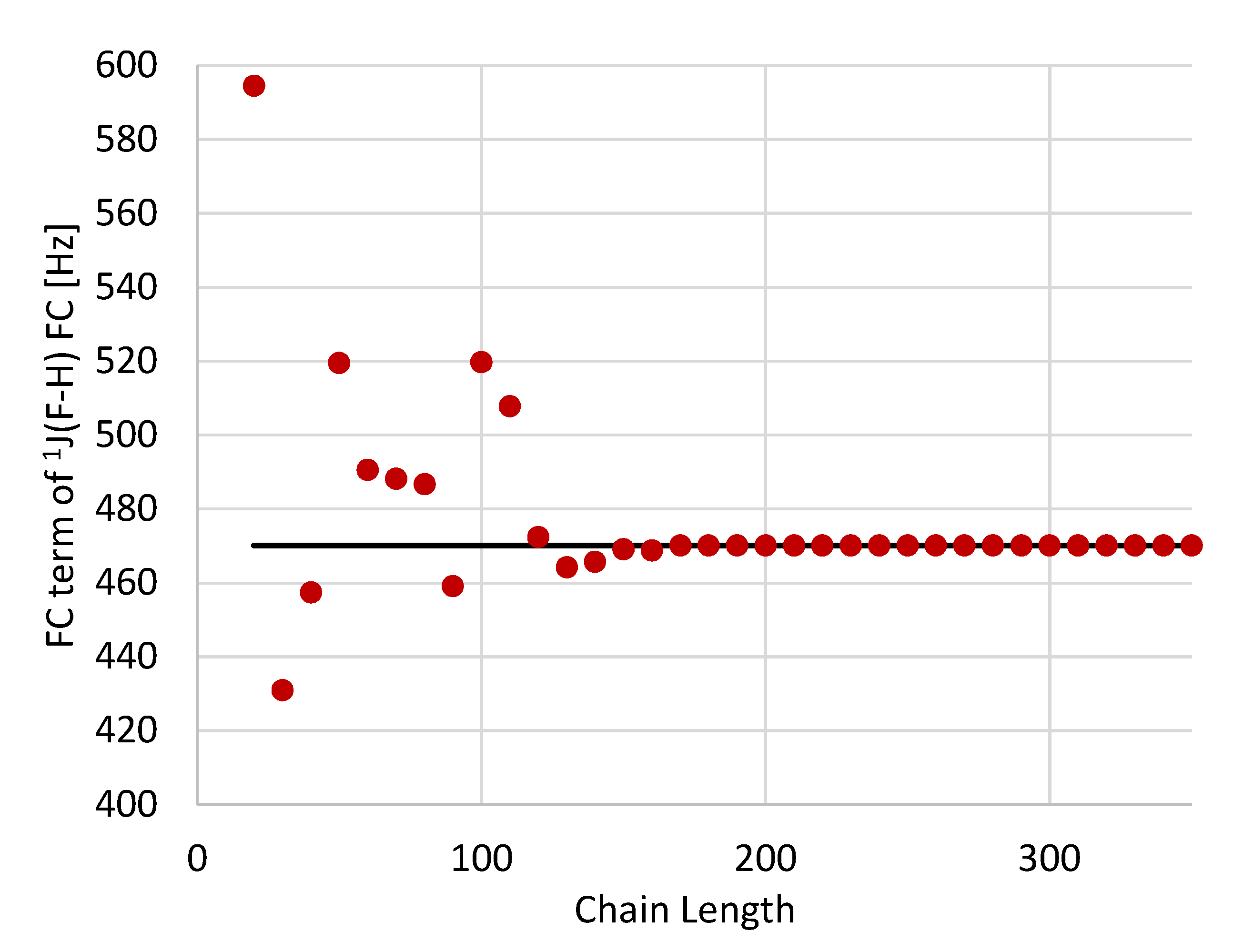}
\caption{HF: The Fermi-contact term for the H-F one-bond coupling, $^1J^{\textrm{FC}}$(F-H), as a function of the Lanczos chain length. The calculations were carried out with the property gradient for the FC operator of F as start vector. The value of the FC term calculated as linear response function is shown as solid line.}
\label{fig:haloF}
\end{figure}

\begin{figure}[!htb]
\centering
  \includegraphics[width=1\linewidth]{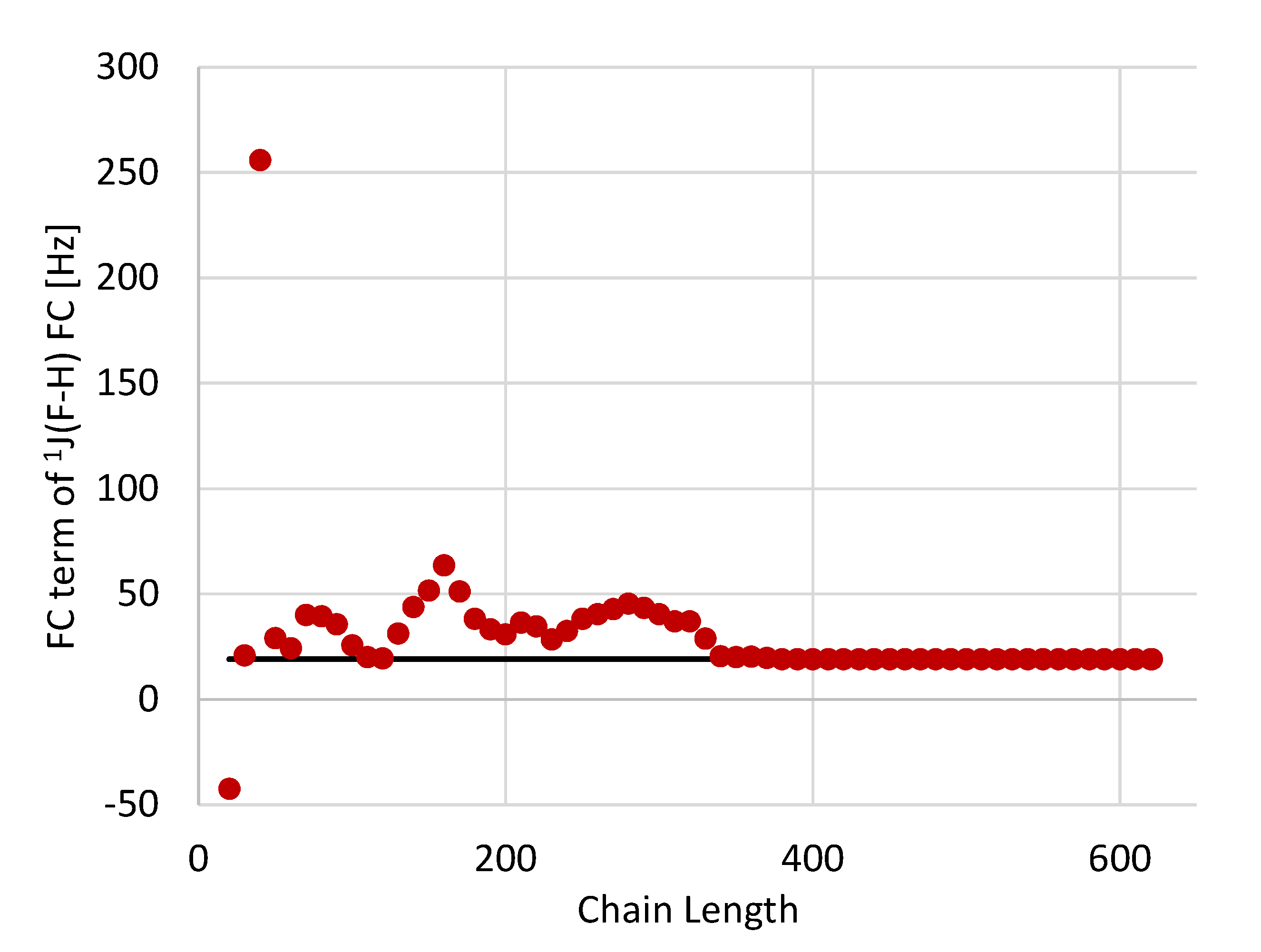}
\caption{HCl: The Fermi-contact term for the H-Cl one-bond coupling, $^1J^{\textrm{FC}}$(Cl-H), as a function of the Lanczos chain length.
The calculations were carried out with the property gradient for the FC operator of Cl as start vector. The value of the FC term calculated as linear response function is shown as solid line.}
\label{fig:H2S-HF-HCl}
\end{figure}

For most coupling constants in this study, the algorithm starts out by underestimating the FC contribution to the coupling constants, as seen for acetylene in Figures \ref{fig:C2H2_H1-C1} and \ref{fig:C2H2_H2-C2}. 
Yet, a few couplings do not follow this most typical convergence pattern for one-bond couplings. 
An example of this is the 
H-O couplings in H$_2$O shown in Figure \ref{fig:H2O_H1-O1}, where the FC term converges with some oscillations from above. Other examples are the H-S couplings in H$_2$S (Figure \ref{fig:H2S_H1-S1}), the H-F coupling in HF (Figure \ref{fig:haloF}) and the H-Cl coupling in HCl (Figure \ref{fig:H2S-HF-HCl}), where the FC terms start with large oscillations around more or less the final value before settling down.

With respect to the number excitations necessary for reaching our 0.5 Hz convergence criterion, in H$_2$O it is 38\% of all the excitations, and 48\% for H$_2$S. Similarly it is 49 \% for HF and 61\%  for HCl. 
Thus, the two hydrides containing third row atoms require a larger percentage of excitations to converge to the desired convergence criterion.

\begin{figure}[!htb]
\centering
  \includegraphics[width=1\linewidth]{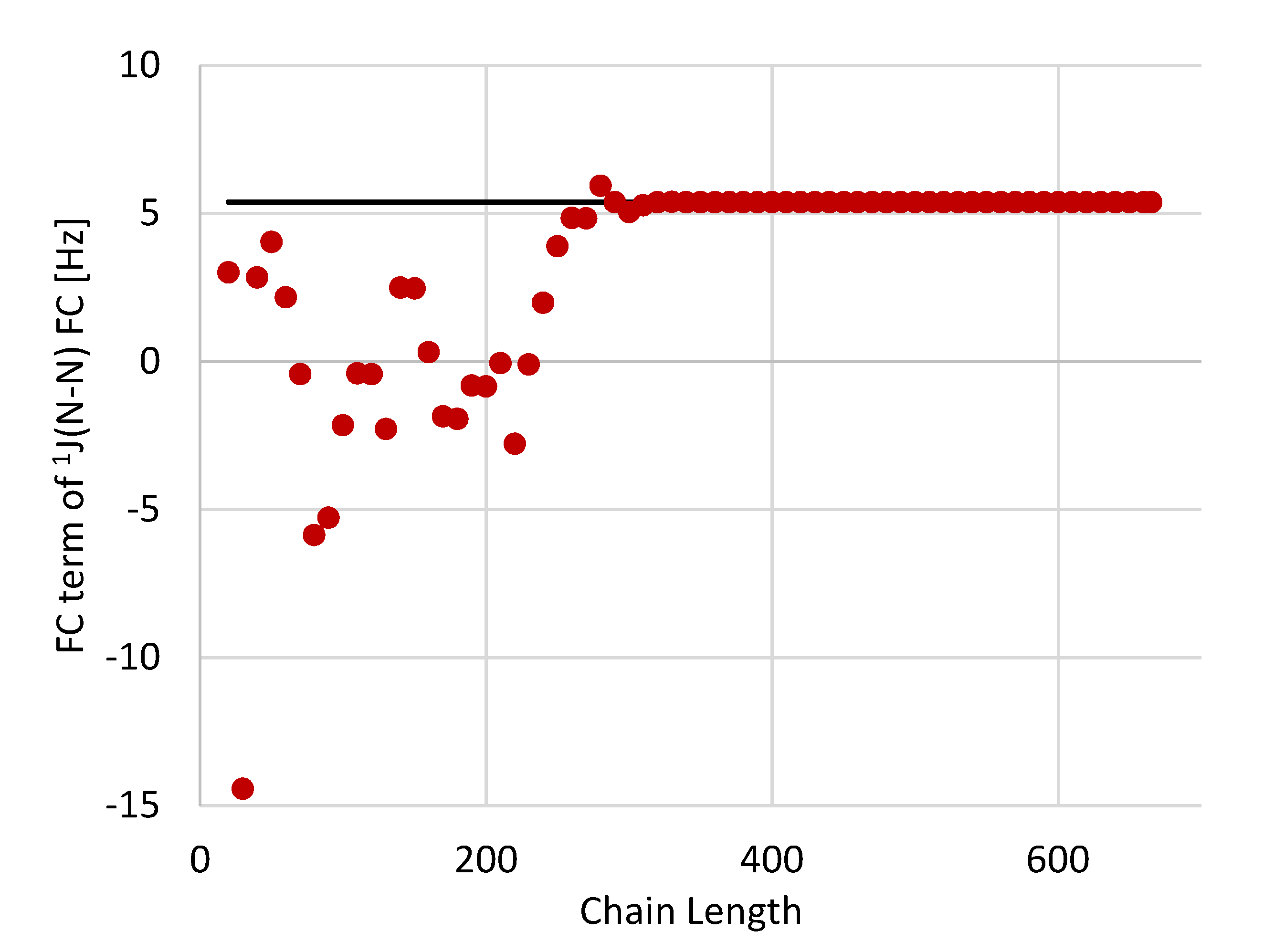}
\caption{N$_2$: The Fermi-contact term for the one-bond N-N coupling, $^1J^{\textrm{FC}}$(N-N), as a function of the Lanczos chain length.
The value of the FC term calculated as linear response function is shown as solid line.
}
\label{fig:N2}
\end{figure}

Turning to the X-Y one-bond couplings, where X and Y are C, N, or O, we observe in Figure \ref{fig:N2} that for the N-N coupling in N$_2$ the FC term converges quite fast with the number of Lanczos vectors. Only about 42\% of the excitation energies are necessary for reaching the allowed deviation of 0.5 Hz.

\begin{figure}[!htb]
\centering
  \centering
  \includegraphics[width=1\linewidth]{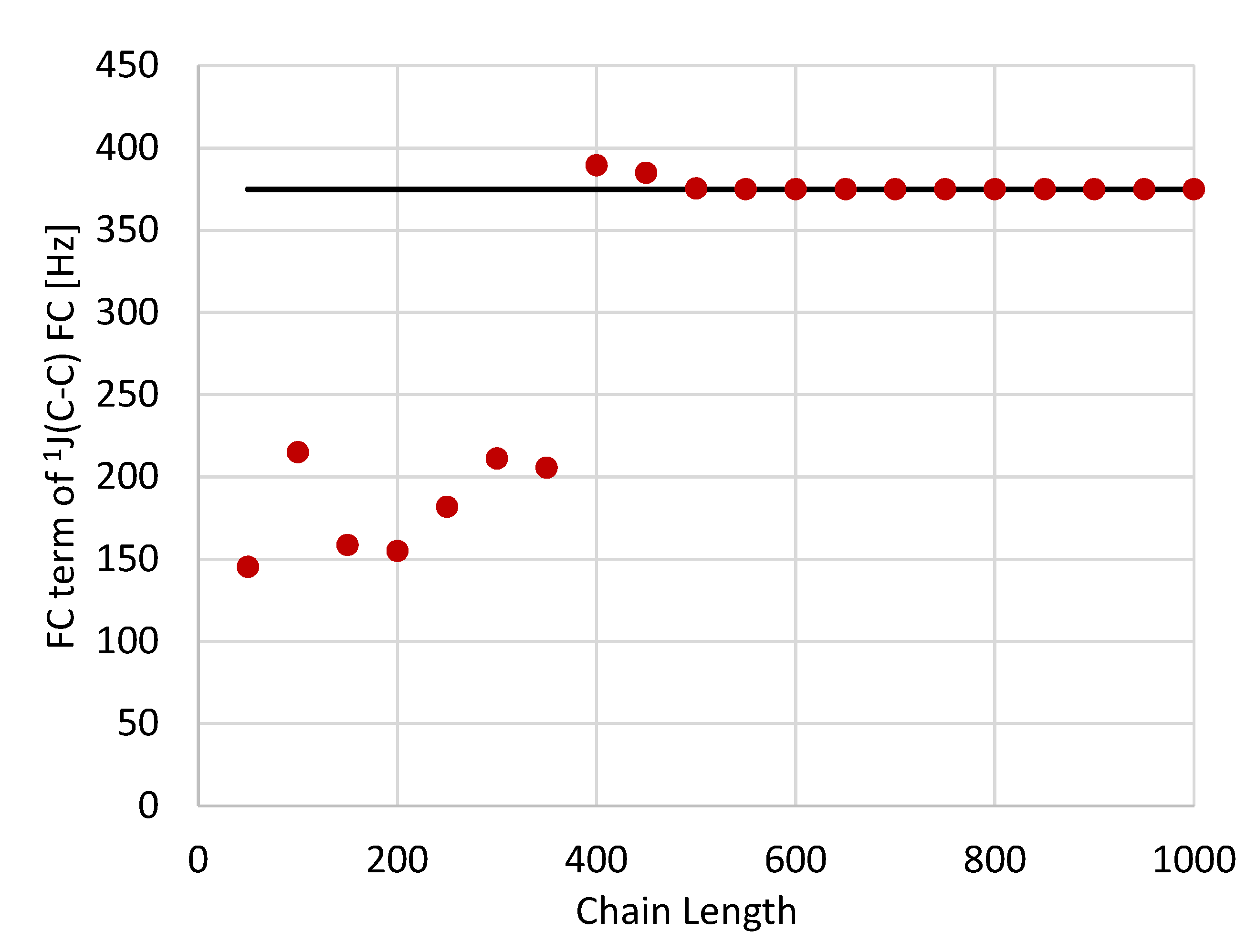}
  \centering
  \includegraphics[width=1\linewidth]{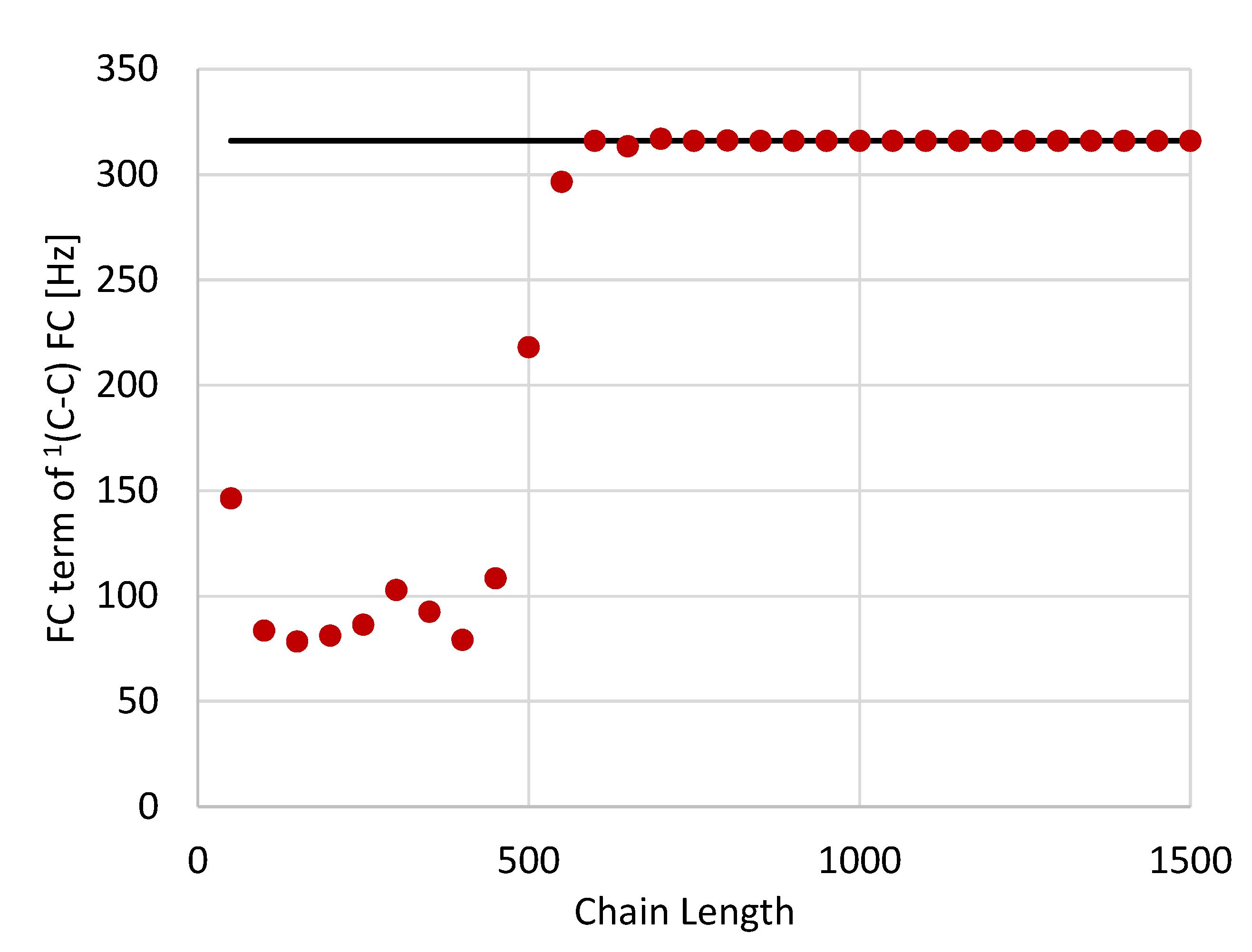}
  \centering
  \includegraphics[width=1\linewidth]{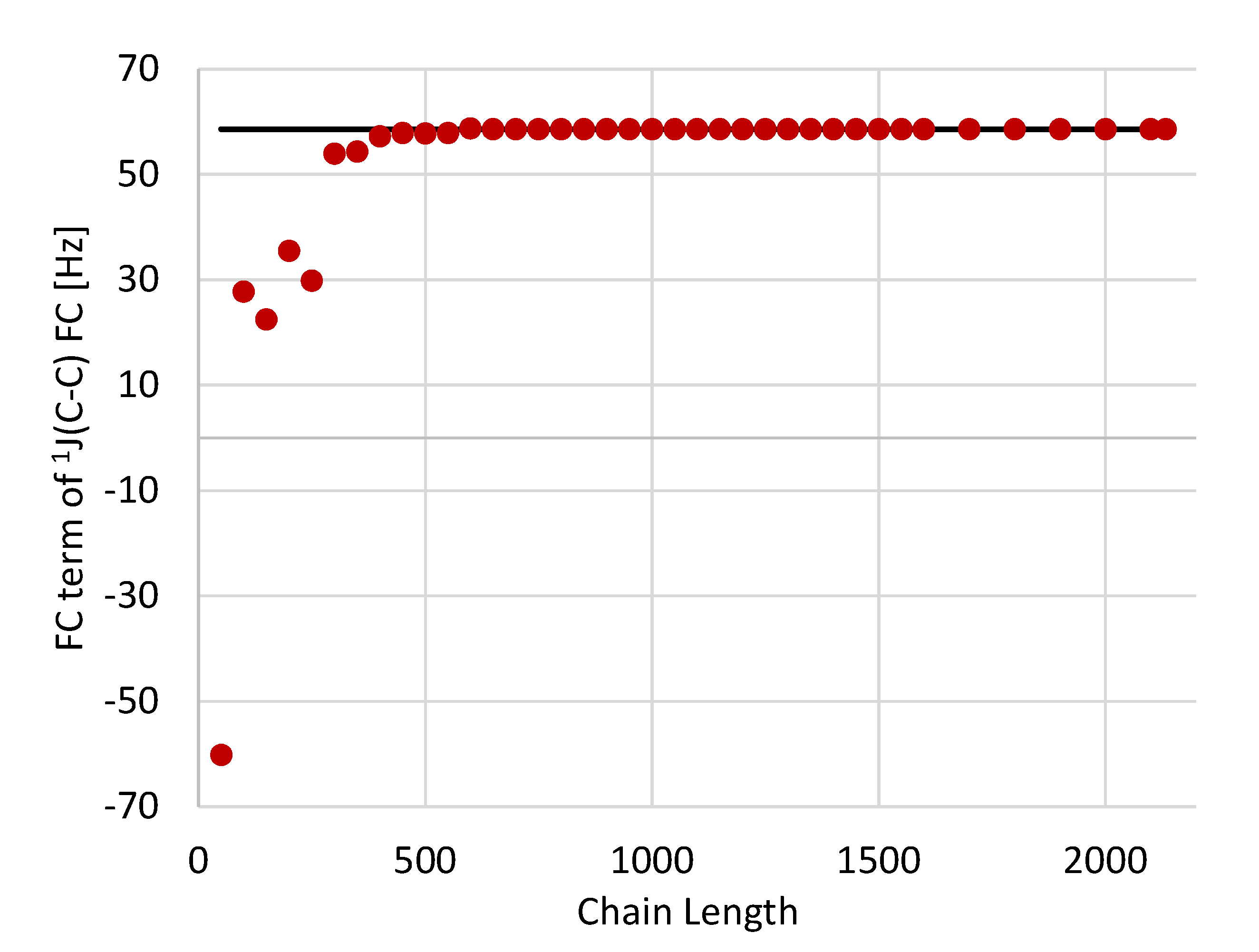}
\caption{The Fermi-contact term for the C-C one-bond coupling, $^1J^{\textrm{FC}}$(C-C), as a function of the Lanczos chain length in \textbf{Top: C$_2$H$_2$ }  \textbf{Center: C$_2$H$_4$} and \textbf{Bottom: C$_2$H$_6$}.
The value of the FC terms calculated as linear response functions are shown as solid lines.}
\label{fig:C-C}
\end{figure}

For the one-bond C-C couplings in acetylene, ethene and ethane, we observe in Figure \ref{fig:C-C} that their FC contributions also converge quite fast, i.e. with a Lanczos chain length between 550 and 750. In percentage of the maximum possible numbers of excitations, which increases from 1001 in acetylene to 2133 in ethane, this corresponds to 55\% for acetylene, 49\% for ethene and only 28\% for ethane.

For the remaining one-bond X-Y couplings between two non-hydrogen atoms, $^1J$(C-O) in CO and H$_2$CO, $^1J$(C-N) in HCN and $^1J$(N-N) as well as $^1J$(N-O) in N$_2$O, we found very similar convergence patterns as for $^1J$(C-C) in acetytlene or ethene. They converge to a maximum deviation of 0.5 Hz between 48\% of the excitation energies (CO and N$_2$O) and 53\% (H$_2$CO).

It is worthwhile to compare the results for the one-bond couplings with the corresponding results obtained by Zarycz et al. \cite{spas153} with the Davidson algorithm, i.e. by summing starting from the lowest the excited state.
They found for the $^1J$(X-H) coupling constants in a subset of the molecules studied here, i.e. CH$_4$, NH$_3$, H$_2$O, SiH$_4$, PH$_3$, H$_2$S, C$_2$H$_2$, C$_2$H$_4$ and C$_2$H$_6$, and for the $^1J$(C-C) couplings in C$_2$H$_2$, C$_2$H$_4$ and C$_2$H$_6$, an approximate functional dependence, which they could fit to a $\tanh(C n)$ dependence of the number of excited states $n$ included. 
However, guaranteed convergence to a constant value  was only reached, when all excited states were included. 
Furthermore, individual very high energy excited pseudo-states could still make large contributions to the coupling constants, leading to large jumps in the results of the partial summations. It is in contrast to the behavior we observe here in Figures \ref{fig:C2H2_H1-C1} to \ref{fig:C-C}.
This implies that the Lanczos algorithm produces much faster approximations to the states, that are important for the one-bond coupling constants.

\subsection{Two-bond couplings}
There are two different types of two-bond couplings: ($i$) the regular ones where atoms that are two bonds apart couple to each other, and ($ii$) the geminal couplings between two hydrogen nuclei, which will be discussed later.
In this study, there are H-C couplings that are two bonds apart in C$_2$H$_2$, C$_2$H$_4$ and C$_2$H$_6$, 
H-O in H$_2$CO, H-N in HCN, and N-O in N$_2$O. 
These couplings lie within a numerical range of 12 Hz (H-C in C$_2$H$_6$) and 240 Hz (N-O in N$_2$O).
For all two-bond couplings, the most dominating term is again the FC term, but for many of these couplings the PSO and SD term also have a considerable impact on the total coupling constant.
Nevertheless, we will again only look at the FC term again. 

\begin{figure}[!h]
    \centering
\includegraphics[width=0.9\linewidth]{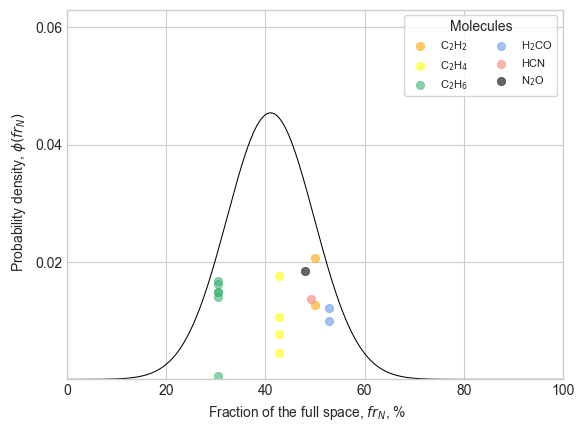}
\caption{The fraction fr$_N$ of the full space of excitations at which the Fermi-contact terms converges with an allowed deviation of 0.5 Hz for all two-bond couplings.}
\label{fig:two-bond_fc_hz}
\end{figure}

In Figure \ref{fig:two-bond_fc_hz}, the fraction of full space needed to converge the FC term is shown for all the two-bond couplings.
The average fraction of the full space needed for the FC term to converge with an allowed deviation of 0.5 Hz is approximately 42$\%$, which is the same as for the one-bond couplings. The C-H two-bond couplings in ethane only need 23.4$\%$ of the full space to converge, whereas the O-H coupling constants in formaldehyde need 52.8$\%$ of the full space. The spread is, therefore, smaller than for the one-bond coupling constants, the number of two-bond couplings is also almost $1/4$ of the number of one-bond coupling constants. 

As for the one-bond coupling constants, the calculations were carried out with FC property gradients as start vectors which correspond to the different couplings. 
The two-bond N-H coupling in HCN behaves differently than all other two-bond couplings, as the FC property gradient for hydrogen as start vector actually gives a better convergence rate  (converged at 49\%) than the nitrogen FC property gradient (53\%). 
For the other two-bond couplings, the FC property gradients for the heavier atoms give the fastest convergence.

In acetylene 
two equivalent two-bond C-H couplings are present, as there are two one-bond C-H couplings. 
Just like the two one-bond couplings, the two two-bond couplings do not converge at the same rate, if one uses the FC property gradient of the same nucleus as start vector for both calculations. 
However, if one uses the corresponding FC property gradients, i.e., the C2 FC property gradient for the C2-H1 coupling and the C1 FC property gradient for the C1-H2 coupling, both couplings show the same convergence pattern.
\begin{figure}[!htb]
\centering
  \includegraphics[width=1\linewidth]{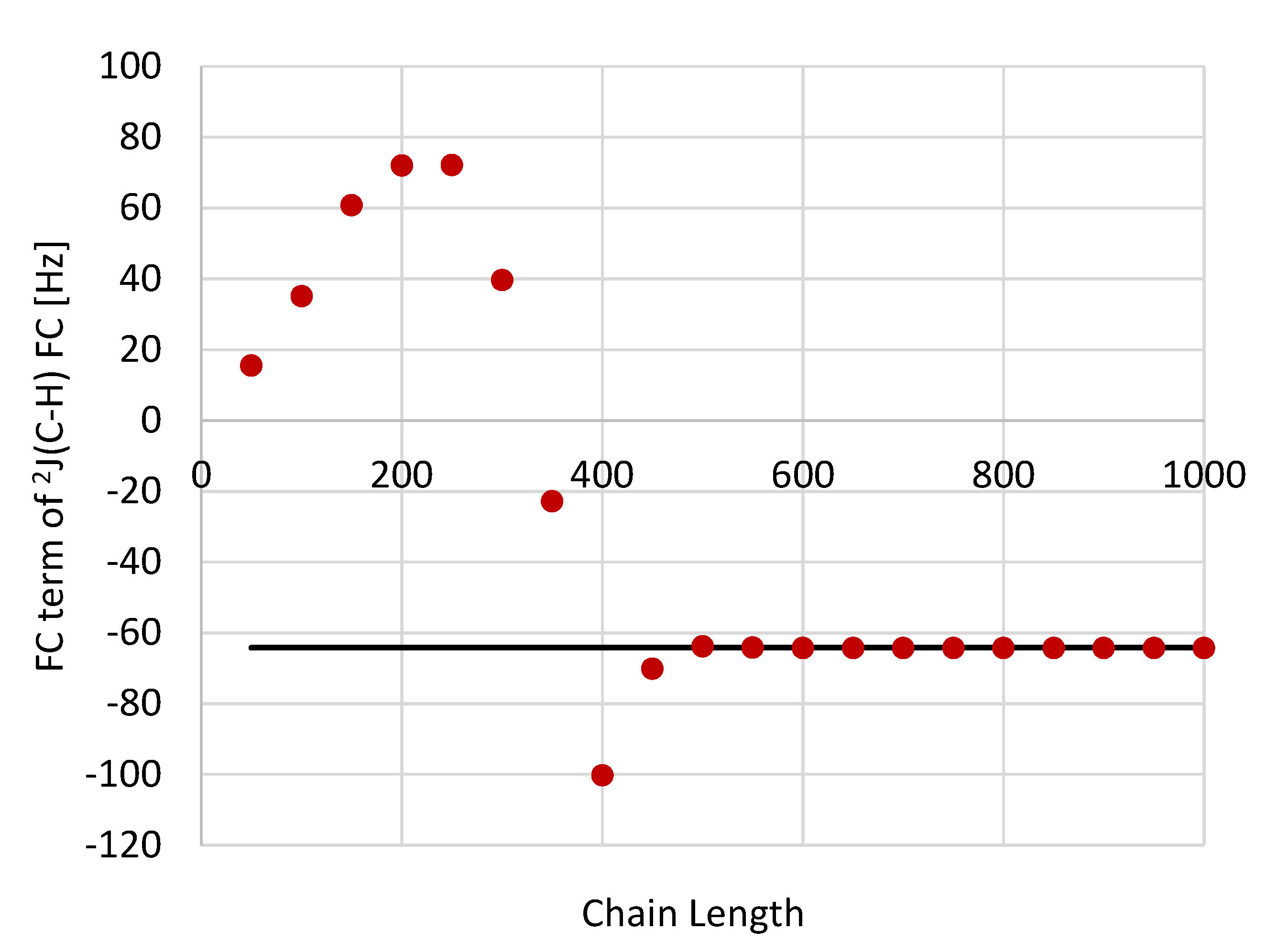}
\caption{C$_2$H$_2$: The Fermi-contact term for the two-bond coupling
between C2 and H1, $^2J^{\textrm{FC}}$(C2-H1), as a function of the Lanczos chain
length. The calculations were carried out with the property gradient
for the FC operator of C2 as start vector.
The value of the FC term calculated as linear response function is shown as solid line.}
\label{fig:C2H2_H1-C2+H2-C1}
\end{figure}
For both ethene and ethane, a similar behaviour as for acetylene is observed. For both molecules, it is important to use a property gradient related to the C atom involved in the coupling as starting vector.

The convergence of the two remaining two-bond couplings, $^2J$(O-H) in formaldehyde and $^2J$(N-O) are very similar to the two-bond C-H couplings in the C$_2$H$_n$ molecules. The FC terms converge at 54\% or 48\% of the excited states.

The results for the $^2J$(C-H) couplings in the C$_2$H$_n$ molecules can again be compared to the corresponding results from the study of Zarycz et al.\cite{spas153} Whereas in their study even very high-lying excited pseudo-states could still lead to spikes in the curves, we observe in our study that the values for the Fermi contact, when converged at a certain number of excited states stay converged when the remaining excited pseudo-state are included. Furthemore, in both studies the coupling converges at lower percentages of excited states for the more saturated compounds.

\subsection{Geminal couplings}
\begin{figure}[!b]
  \includegraphics[width=1\linewidth]{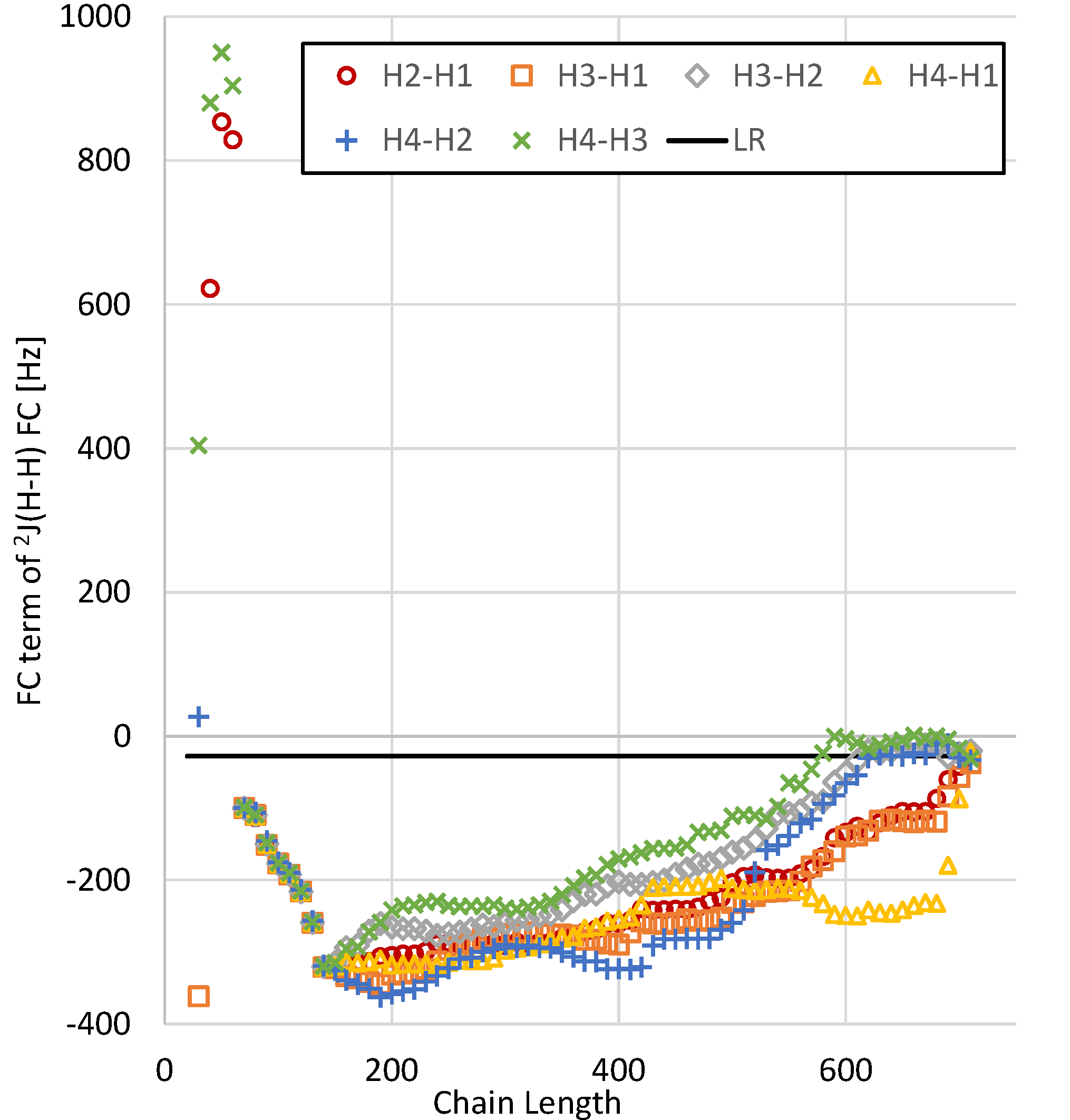}
\caption{CH$_4$: 
The Fermi-contact term for the various geminal H-H coupling constants, $^2J^{\textrm{FC}}$(H-H), as a function of the Lanczos chain
length.
The calculations were carried out with the property gradient
for the FC operator of the carbon atom as start vector.
The value of the FC term calculated as linear response function is shown as solid line.
}
\label{fig:methane-C}
\end{figure}
In $^1$H NMR the term geminal coupling refers to a coupling between two hydrogen atoms on the same carbon atom, i.e. H-C-H. 
In this study, this term will also be applied to couplings between two hydrogen atoms on atoms different from carbon, i.e. H-X-H where X = B, N, O, Si, P, and S. 
The geminal coupling constants numerically range from 1.9 Hz (SiH$_4$) to 90.5 Hz (C$_2$H$_4$). 
All the coupling constants are negative except for the one for the geminal coupling in formaldehyde which is 16.6 Hz.
As for the other types of couplings, the FC term is for the geminal couplings the most dominant one, when it comes to the total coupling constant. 

For the one- and two-bond couplings, the X-H coupling constants for all hydrogen atoms placed on the same atom followed the same convergence pattern, e.g., all the B-H coupling constants in BH$_3$ or all the C-H in C$_2$H$_6$. 
For the geminal couplings, this is not necessarily the case. This is demonstrated in Figure \ref{fig:methane-C} for the geminal coupling constants in methane, where the convergence plots for all six equivalent H-H couplings in methane are shown, using the FC property gradient vector of the carbon atom as start vector. The only thing they all have in common is that they do not converge until the very end. 
The six equivalent couplings do not even reach the same final value for either of the total coupling constants, ranging from $-19.97$ Hz to $-37.32$ Hz, and therefore they do also not reach the same value as obtained in the linear response calculation, $-27.67$ Hz. 
However, if the mean value for all the equivalent couplings is taken, the value is the same as the linear response result. 
\begin{figure}[!h]
  \includegraphics[width=1\linewidth]{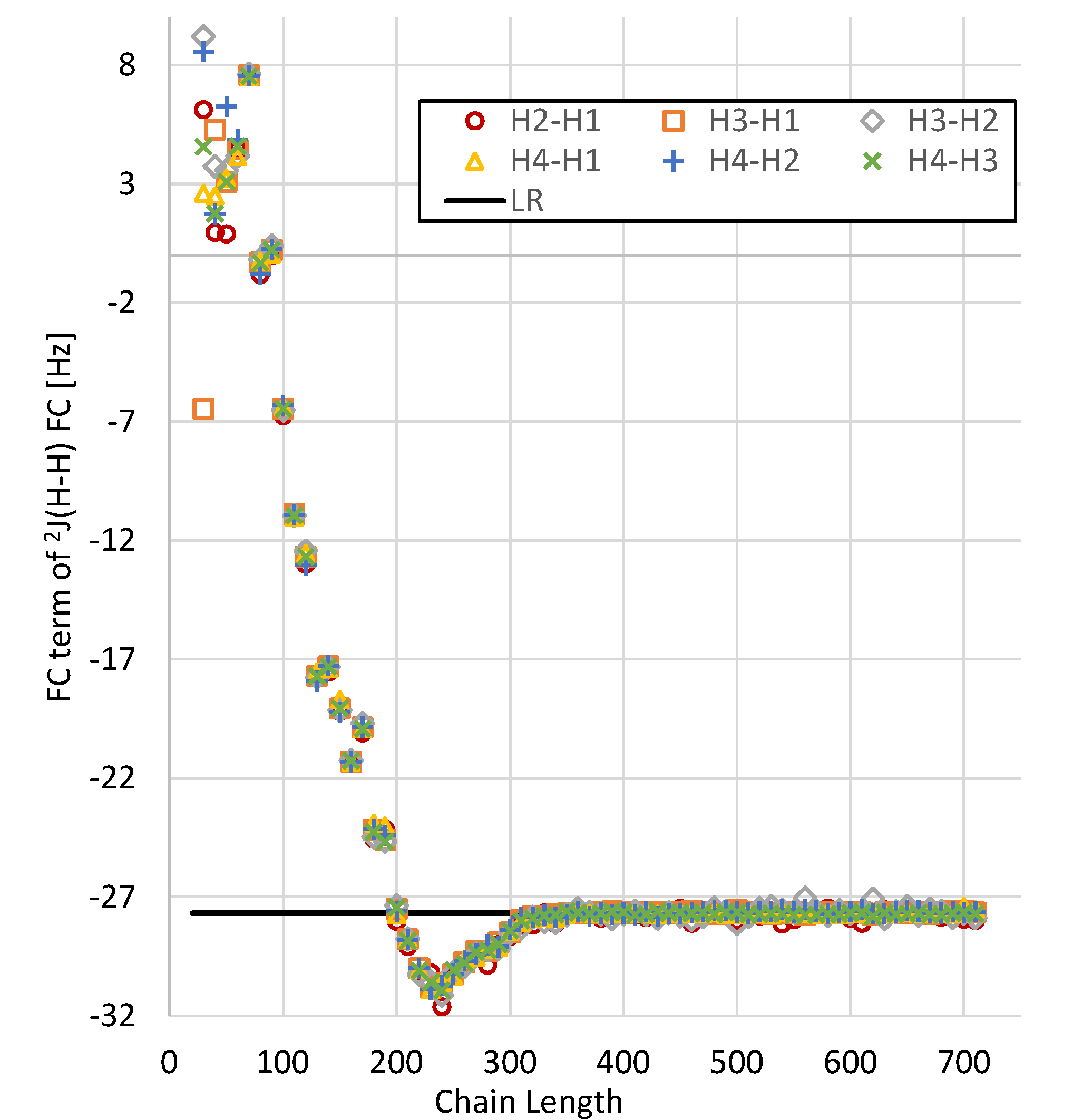}
\caption{CH$_4$: 
The Fermi-contact term for the various geminal H-H coupling constants, $^2J^{\textrm{FC}}$(H-H), as a function of the Lanczos chain
length.
The calculations were carried out with the property gradient
for the FC operator of one of the two coupled hydrogen atoms as start vector.
The value of the FC term calculated as linear response function is shown as solid line.
}
\label{fig:methane}
\end{figure}

For the one-bond couplings it was observed that choosing a FC property gradient corresponding to one of the coupled atoms as starting vector made the FC term converge faster. For the molecules with equivalent geminal couplings this does help also as shown in Figure \ref{fig:methane} again for the geminal coupling constants in methane, where for each of the couplings the FC property gradient of one of the coupled atoms was now used as start vector. 
Changing which of the two property gradients is used does not change the results significantly.
Now all the geminal coupling constants converge to the linear response result in the same way and already with only 44\% of the excited pseudo-states included, which is comparable to the other types of couplings. 
The same trend can be observed for all other molecules with equivalent geminal couplings. 

\begin{figure}[!htb]
\centering
  \includegraphics[width=1\linewidth]{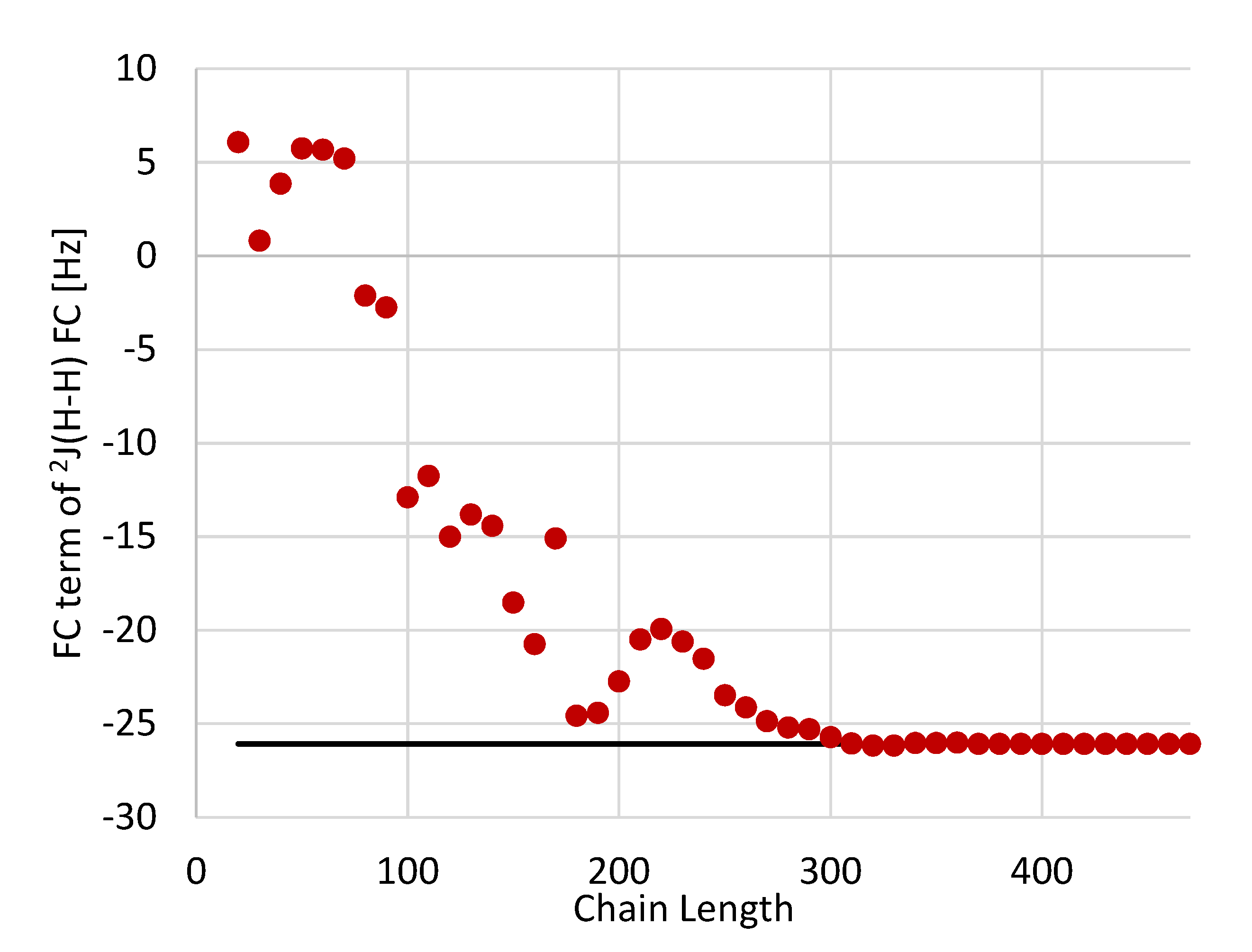}
\caption{H$_2$O: The Fermi-contact term for the geminal H-H coupling, $^2J^{\textrm{FC}}$(H-H), as function of the Lanczos chain length. The calculations were carried out with the property gradient for the FC operator of H1 as start vector.
The value of the FC term calculated as linear response function is shown as solid line.}
\label{fig:HH-H2O}
\end{figure}
\begin{figure}[!htb]
\centering
  \includegraphics[width=1\linewidth]{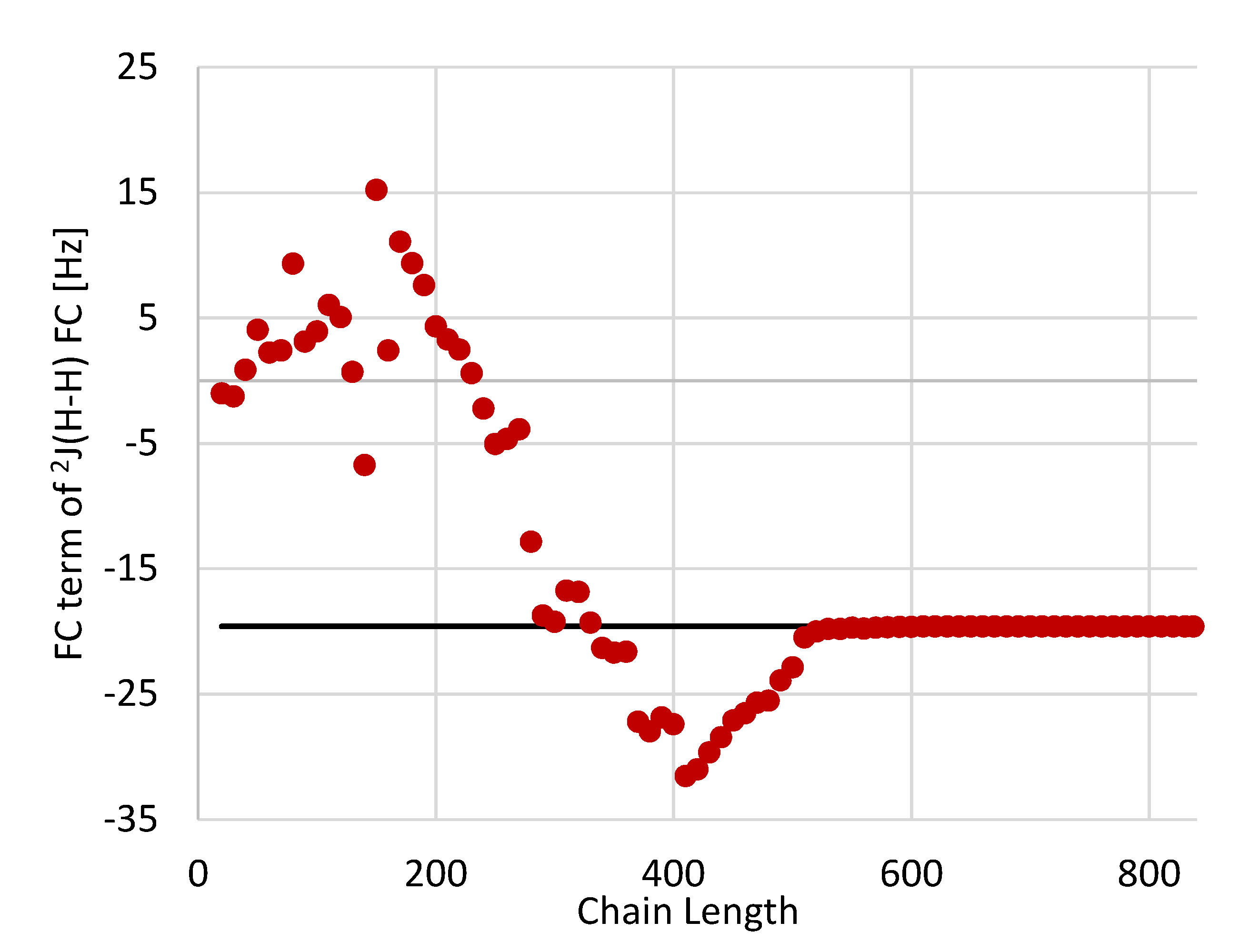}
\caption{H$_2$S: The Fermi-contact term for the geminal H-H coupling, $^2J^{\textrm{FC}}$(H-H), as function of the Lanczos chain length. The calculations were carried out with the property gradient for the FC operator of H1 as start vector.
The value of the FC term calculated as linear response function is shown as solid line.}
\label{fig:HH-H2S}
\end{figure}
\begin{figure}[!htb]
\centering
  \includegraphics[width=1\linewidth]{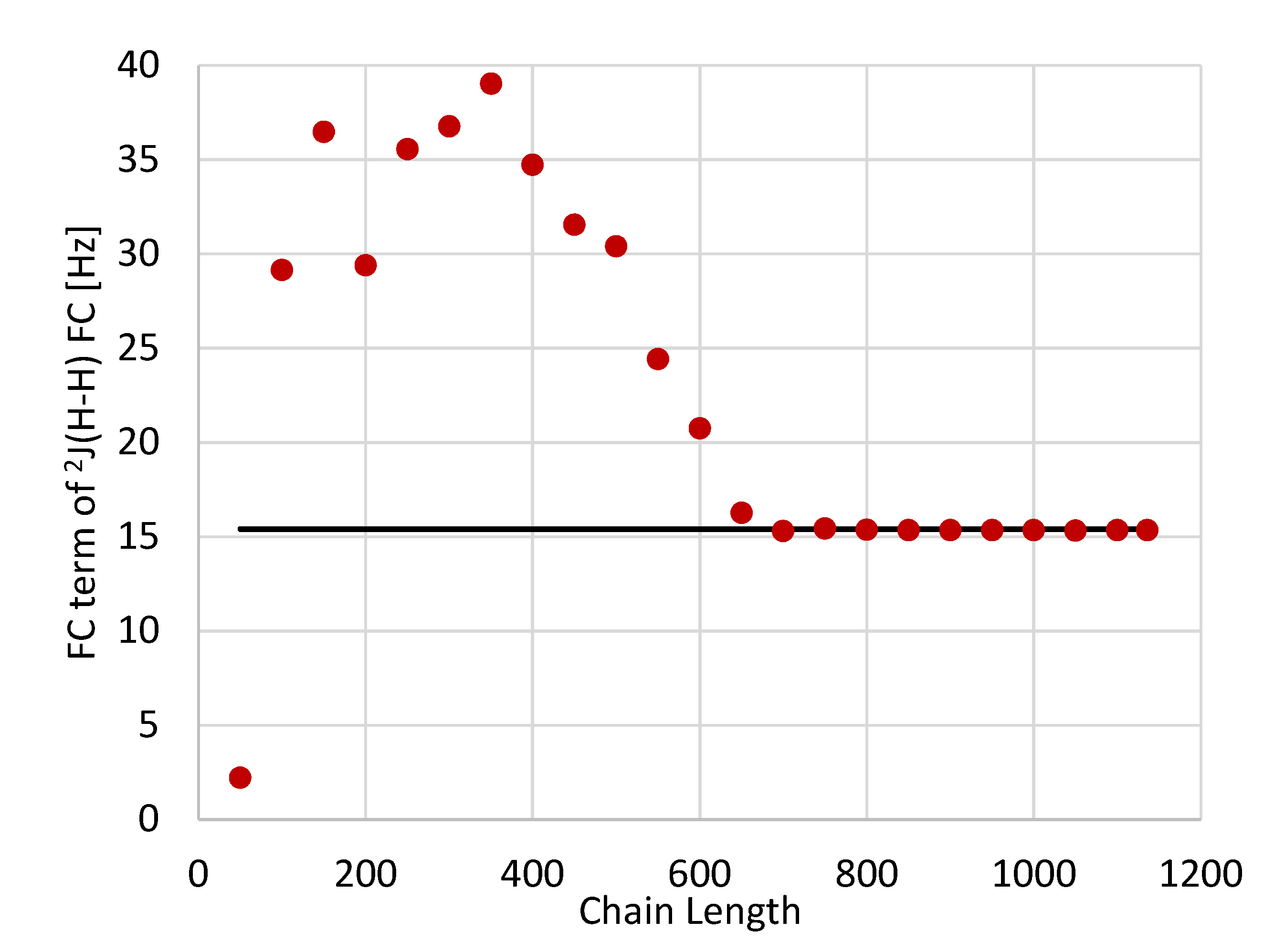}
\caption{H$_2$CO: The Fermi-contact term for the geminal H-H coupling, $^2J^{\textrm{FC}}$(H-H), as function of the Lanczos chain length. The calculations were carried out with the property gradient for the FC operator of H1 as start vector.
The value of the FC term calculated as linear response function is shown as solid line.}
\label{fig:HH-H2CO}
\end{figure}
\begin{figure}[!htb]
\centering
  \includegraphics[width=1\linewidth]{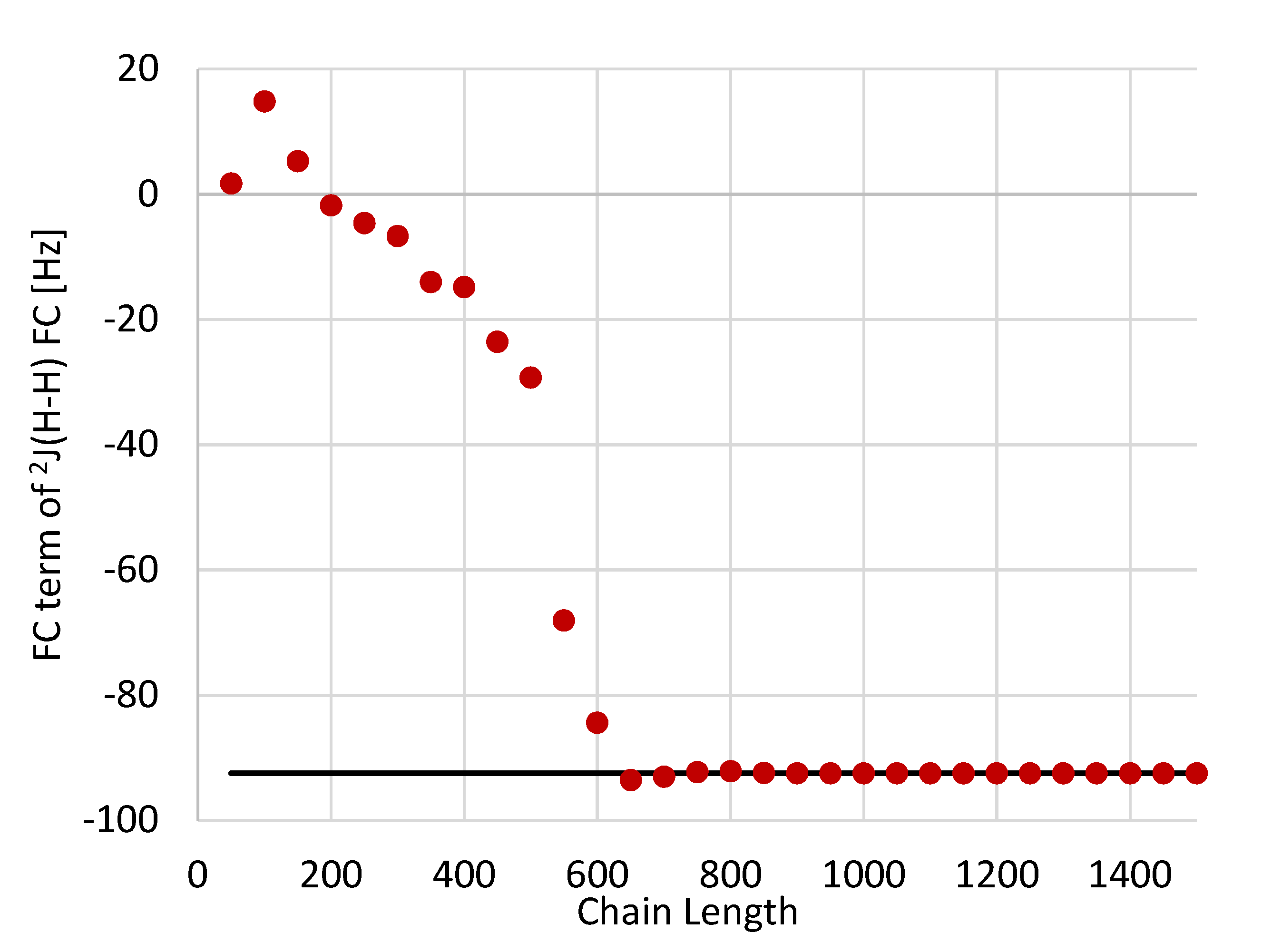}
\caption{C$_2$H$_4$: The Fermi-contact term for the geminal H1-H2 coupling, $^2J^{\textrm{FC}}$(H-H), as function of the Lanczos chain length. The calculations were carried out with the property gradient for the FC operator of H2 as start vector.
The value of the FC term calculated as linear response function is shown as solid line.}
\label{fig:HH-C2H4}
\end{figure}
For the molecules containing only one geminal coupling, choosing 
one of the hydrogen's FC property gradient as start vector also gives good results. The FC terms for the geminal couplings in H$_2$O, H$_2$S, H$_2$CO and C$_2$H$_4$ all converge, as can be observed in Figures \ref{fig:HH-H2O}, \ref{fig:HH-H2S}, \ref{fig:HH-H2CO} and \ref{fig:HH-C2H4}. 
For the first three molecules at 60\% to 62\% of the total number of excited pseudo-states while for C$_2$H$_4$ already 42\% of the excited states suffice.


Comparing again to the results obtained by Zarycz et al.\cite{spas153} for the geminal couplings in CH$_4$, H$_2$O and H$_2$S, Figures \ref{fig:methane}, \ref{fig:HH-H2O} and \ref{fig:HH-H2S}, we observe that in both methods the coupling constants converge to the final value around the same number of exited pseudo-states included in the summation. 
However, in the Davidson algorithm, i.e. summing the excited states from below, Zarycz et al. encountered again spikes in the results of the partial summations even for quite high energy excited pseudo-states, which cannot happen in the Lanczos algorithm.
For ethane, on the other hand, there are large differences between the two algorithms. In the Lanczos approach, Figure \ref{fig:HH-C2H4}, the Fermi contact term already converged at 42\% of the excited states included, while in the Davidson algorithm convergence was not reached until around 75\% and even then spikes were observed for larger numbers of excited states included.

\subsection{Vicinal couplings}
\begin{figure}[!b]
\centering
  \includegraphics[width=1\linewidth]{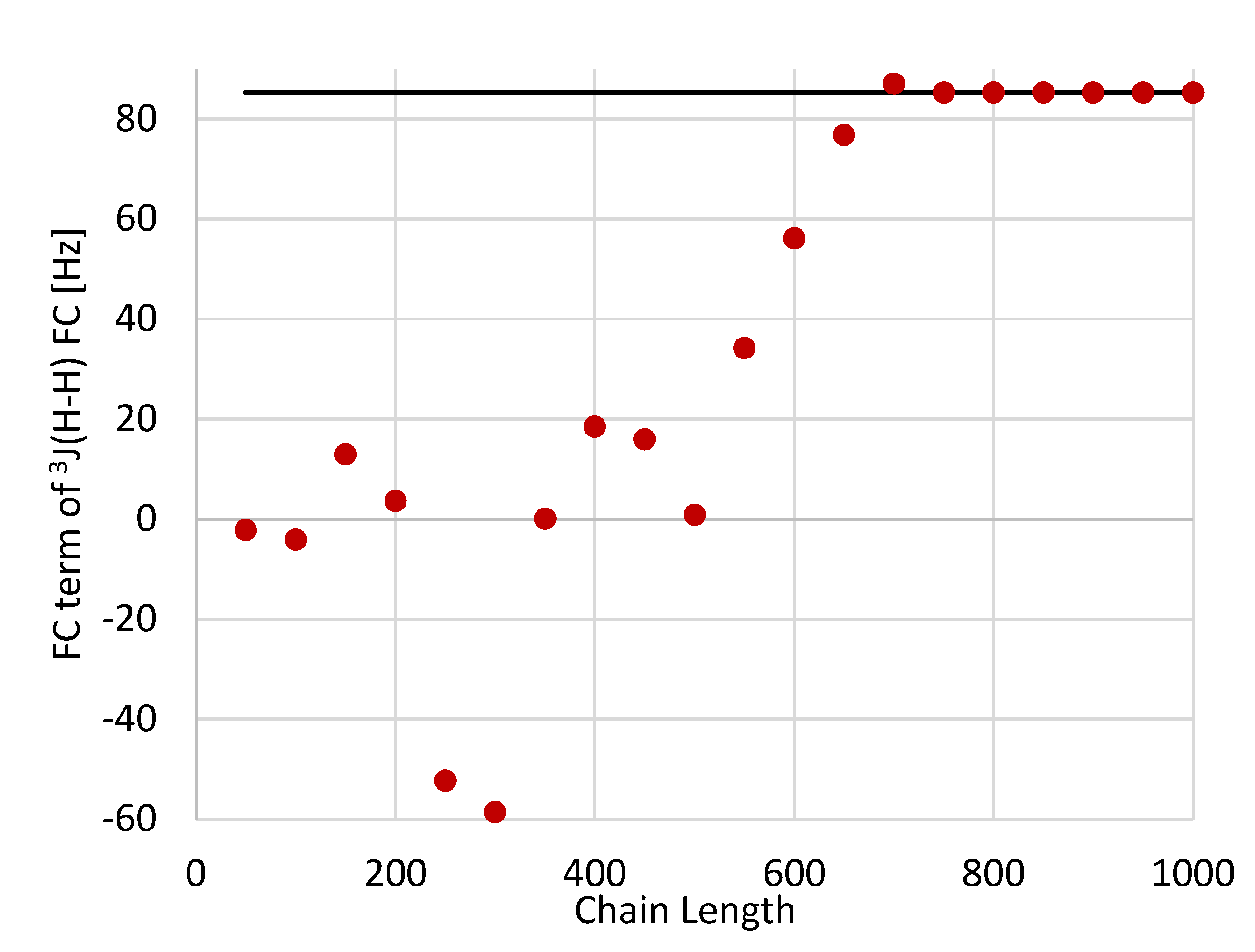}
  \includegraphics[width=1\linewidth]{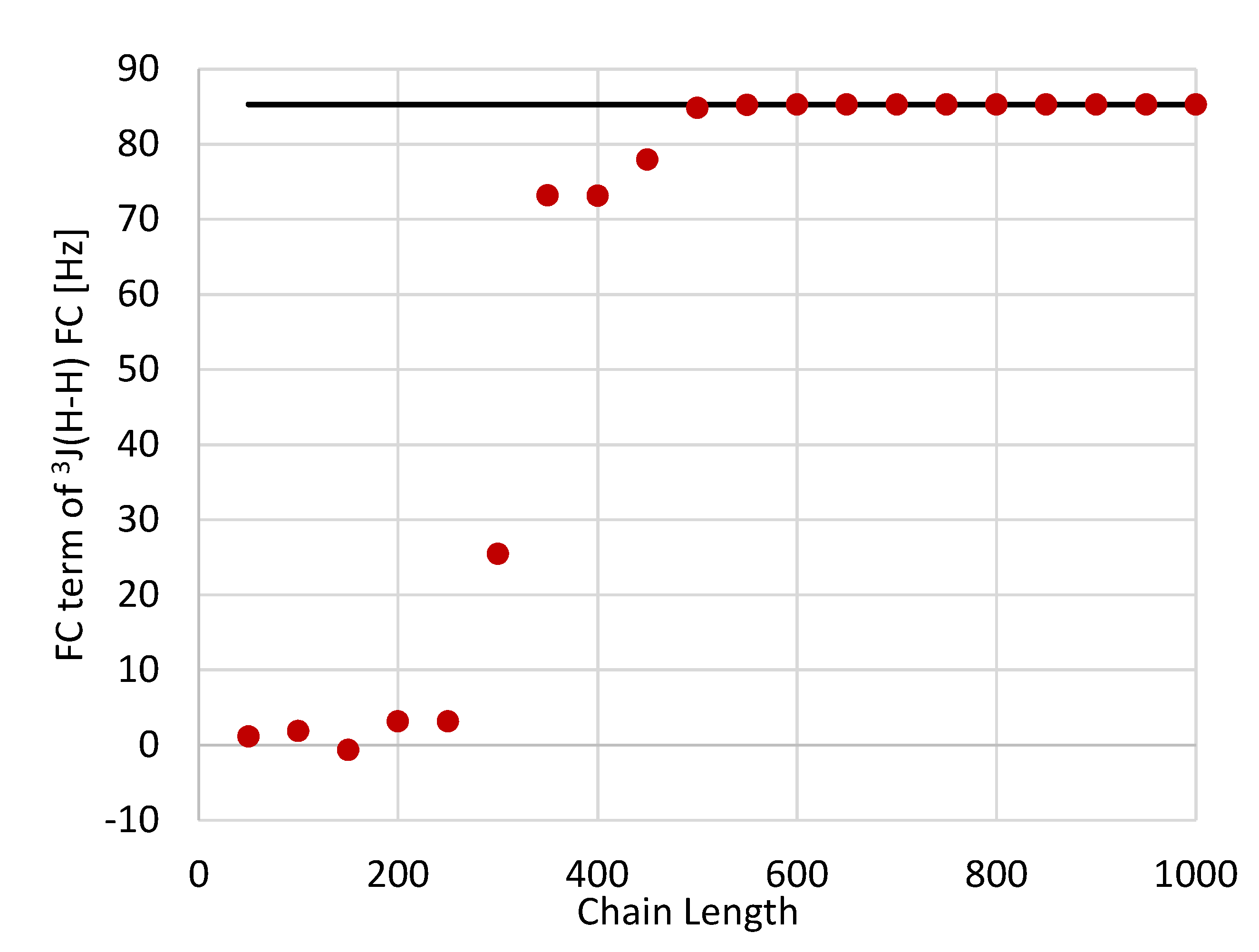}
\caption{C$_2$H$_2$: The Fermi-contact term for the vicinal H-H coupling, $^3J^\textrm{FC}$(H-H), as function of the Lanczos chain length. \textbf{Top:} 
The calculations were carried out with the property gradient for the FC operator of C as start vector.
\textbf{Bottom:} 
The calculations were carried out with the property gradient for the FC operator of H as start vector.
The value of the FC terms calculated as linear response functions are shown as solid lines.}
\label{fig:vic_C2H2}
\end{figure}
A vicinal coupling is like a geminal coupling, only three bonds apart instead of two, which means it is a coupling between two hydrogen atoms on two carbon atoms next to each other. In this study this type of coupling is present in C$_2$H$_2$, C$_2$H$_4$ and C$_2$H$_6$. 
For this type of coupling, as for most but not all of the couplings in this study, the most dominating term is the FC term.  

The vicinal coupling in acetylene converges almost with the same pattern no matter if the FC integral for C1 or C2 is used, Figure \ref{fig:vic_C2H2} above.
However, using the carbon atom FC property gradient as starting vector it does not converge before 75\% of the excited pseudo-states are included. 
Using instead of the FC property gradient for the hydrogen atoms convergence is already achieved at 50\%, as seen in the bottom of Figure \ref{fig:vic_C2H2}.

\begin{figure}[!b]
\centering
  \includegraphics[width=1\linewidth]{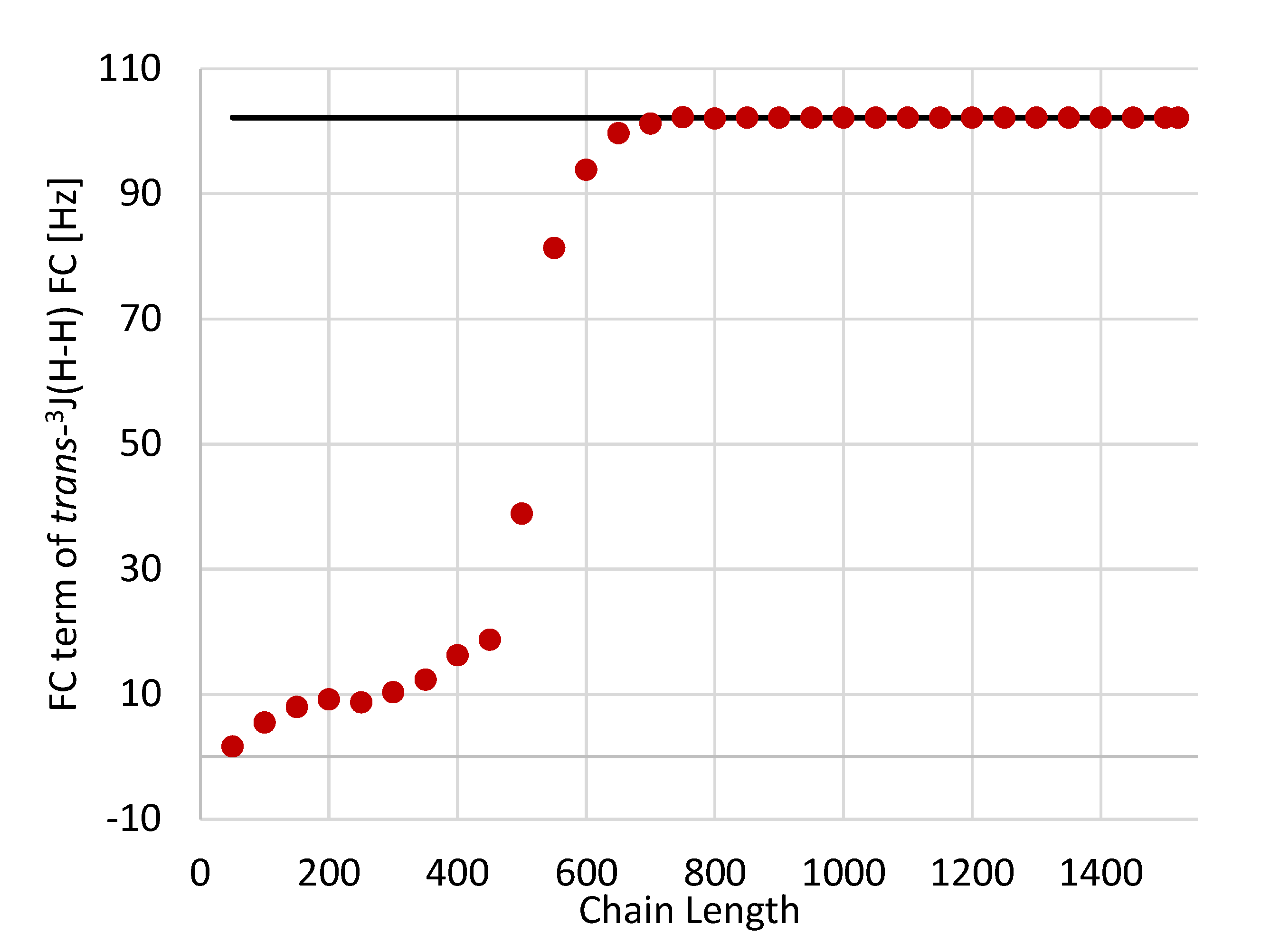}
  \includegraphics[width=1\linewidth]{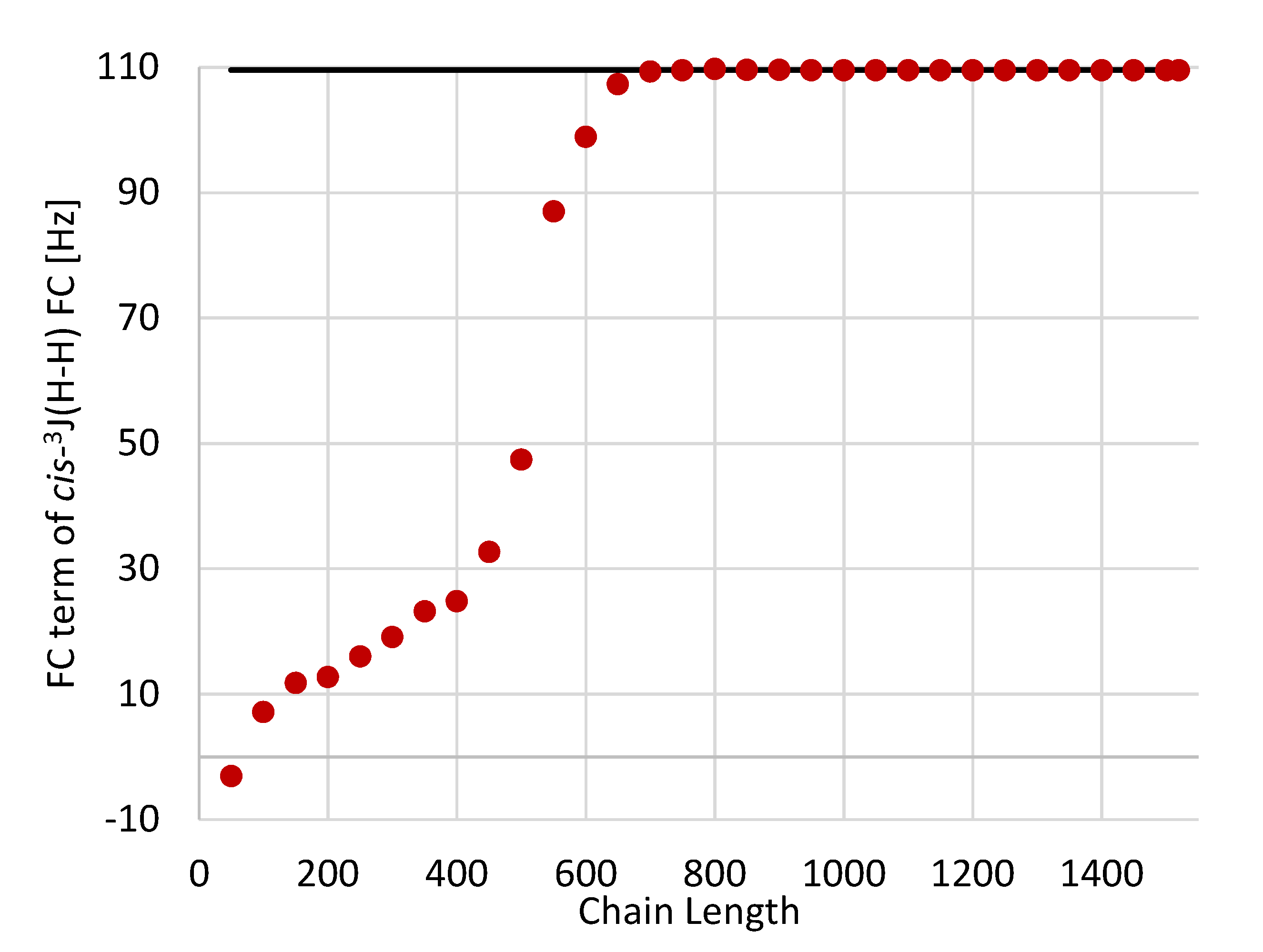}
\caption{C$_2$H$_4$: \textbf{Top:} The Fermi-contact term for the \textit{trans} vicinal H-H coupling, \textit{trans}-$^3J^{\textrm{FC}}$(H-H), as function of the Lanczos chain length. \textbf{Bottom:} The Fermi-contact term for the \textit{cis} vicinal H-H coupling, \textit{cis}-$^3J^{\textrm{FC}}$(H-H), as function of the Lanczos chain length.
Both were calculated the FC property gradient of one of the coupled atoms as start vector
The values of the FC terms calculated as linear response functions are shown as solid lines.}
\label{fig:vic_C2H4}
\end{figure}
For ethene, there are two different types of vicinal couplings. 
Two \textit{cis}-couplings between H3-H2 and H4-H1 and two \textit{trans}-coupling between H3-H1 and H4-H2. 
The Fermi contact terms converge nicely already at 46\% of the excited pseudo-states included for the \textit{cis}-coupling and at 49\% for the \textit{trans}-coupling, when the FC property gradient for one of the coupled hydrogen atoms is used as start vector, as can be seen in Figure \ref{fig:vic_C2H4}.
Both \textit{cis}- and both \textit{trans}-couplings show the same convergence pattern and converged to the same value, which is equal to the linear response result.


\begin{figure}[!b]
\centering
  \includegraphics[width=1\linewidth]{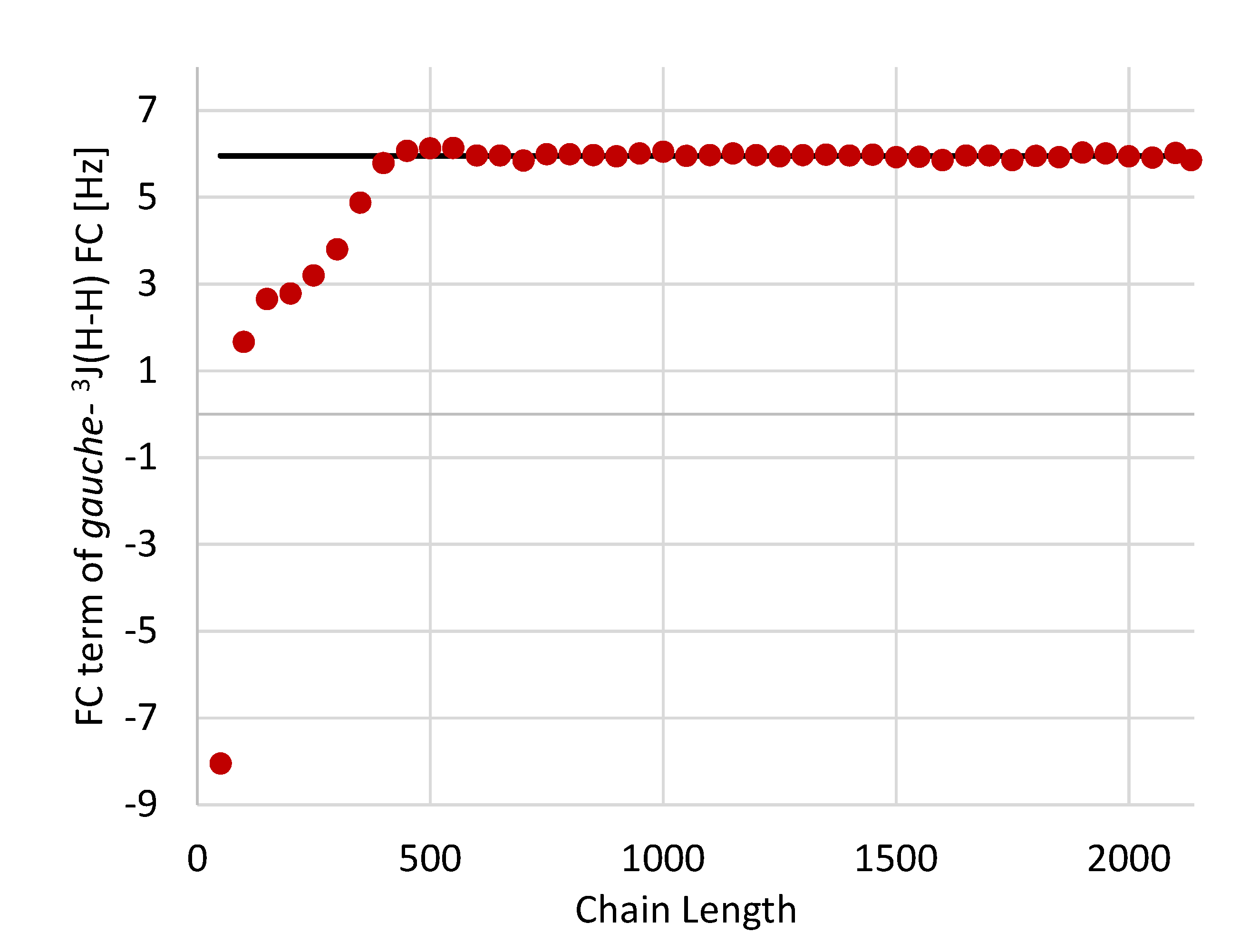}
  \includegraphics[width=1\linewidth]{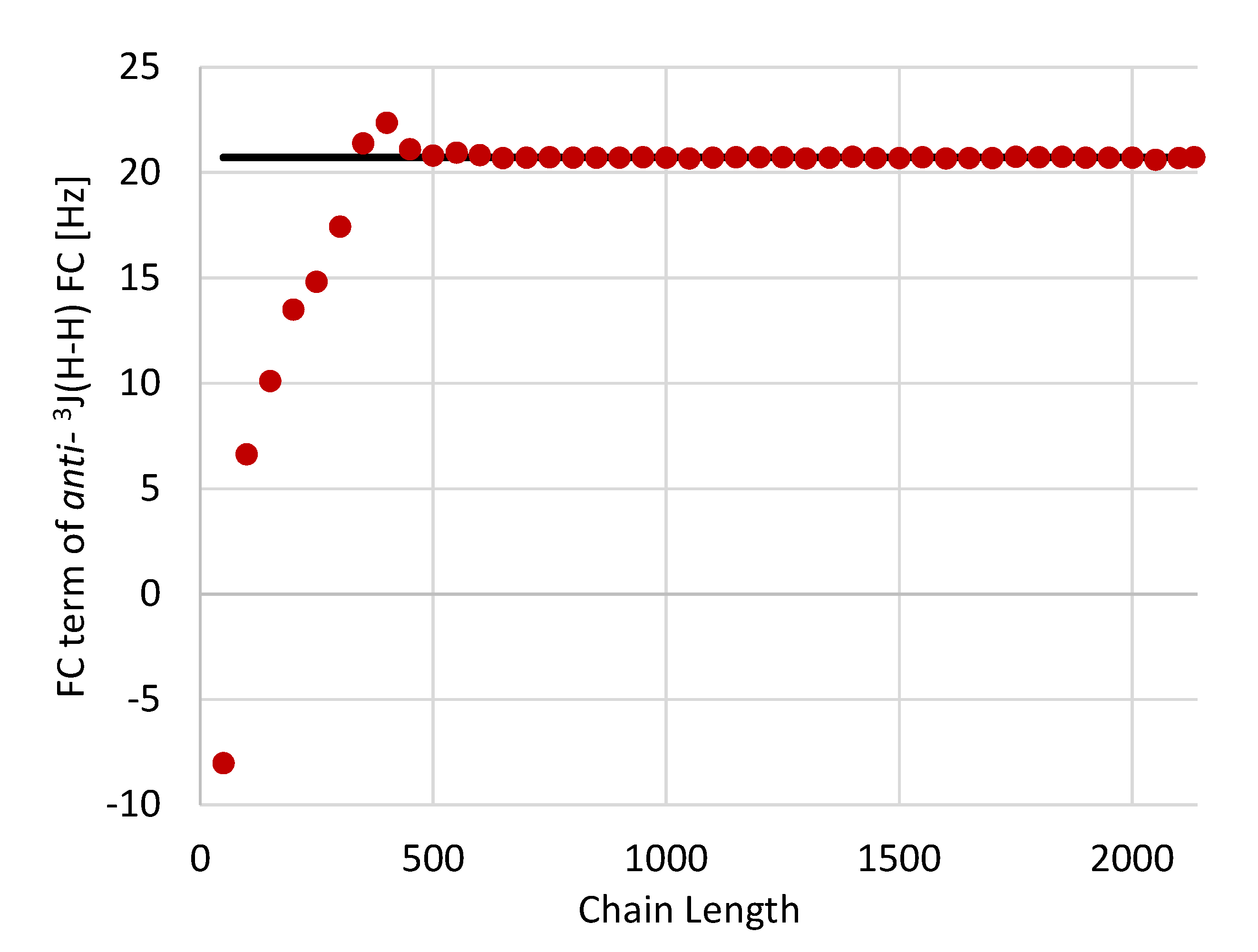}
\caption{C$_2$H$_6$: \textbf{Top:} The Fermi-contact term for the \textit{gauche} vicinal H-H coupling, \textit{gauche}-$^3J^{\textrm{FC}}$(H-H), as function of the Lanczos chain length. \textbf{Bottom:} The Fermi-contact term for the \textit{anti} vicinal H-H coupling, \textit{anti}-$^3J^{\textrm{FC}}$(H-H), as function of the Lanczos chain length. Both were calculated with the FC property gradient of one of the coupled atoms as start vector.
The value of the FC terms calculated as linear response functions are shown as solid lines.}
\label{fig:vic_C2H6}
\end{figure}
For ethane, there would only be one type of vicinal coupling constant in NMR experiments, because the two methyl groups are free to rotate. 
When doing calculations for a fixed geometry, this is, however, not the case.
Therefore there are two different types of vicinal couplings, six \textit{gauche}-coupling constants, and three \textit{anti-periplanar}-coupling constants. 
For both types, the Fermi contact term converges very fast, if they are started with the FC property gradient of one of the coupled atoms as start vector, i.e. for the \textit{gauche}-$^3J$(H-H) at 19\% of the excited pseudo-states included in the summation and for the \textit{anti-periplanar}-$^3J$(H-H) at 21\% as can be seen in Figure \ref{fig:vic_C2H6}.
This holds for all the six equivalent \textit{gauche}-coupling constants, and also for the three equivalent \textit{anti-periplanar}-coupling constants.
The vicinal couplings behave therefore very similar to the geminal couplings. 

The comparison with the work of Zarycz et al.\cite{spas153} shows a similar situation than for the geminal couplings. When e.g. for acetylene or ethene the Fermi contact term in the Lanczos algorithm is converged to the linear response result with between 46\% and 50\% of the excited pseudo-states included and for ethane already at 20\%, it stays so until all excited pseudo-states are included, whereas in the case of the Davidson algorithm oscillations around the linear response result continue until almost all states are included.


\section{Conclusion}
The aim of this study was twofold, namely 
to extend our former implementation of the implicit block Lanczos algorithm\cite{ZamokJCP2021} in the Dalton program to the calculation of NMR SSCCs,
and then to investigate, 
whether it could be used to truncate the number of excited states necessary for the sum-over-states calculation of the often dominant Fermi contact contribution to the SSCCs and still achieve reasonable results. 
The hope was to achieve a more reliable method than with the Davidson algorithm,\cite{spas153} where for the Fermi contact term almost all states had to be included or some extrapolation procedure had to be used.

For almost all the couplings, the algorithm required less than 50$\%$ of the full space in order for the FC term to converge within a 0.5Hz deviation from the linear response results.
The highest percentages of excited states, i.e. 61\%, were only necessary for a few couplings: the one-bond couplings in PH$_3$ and HCl, and the geminal hydrogen-hydrogen couplings in H$_2$O, H$_2$S and H$_2$CO.
The lowest percentages of necessary Lanczos vectors were observed for all the couplings in ethane, with values between 19\% and 28\% corresponding to inbetween 400 and 600 Lanczos vectors.
Unlike in a former study employing the Davidson algorithm,\cite{spas153} when the Lanczos algorithm had produced a converged result for the Fermi contact contribution for a truncated set of excited pseudo-states, it did not oscillate anymore until all excited states were included (full space).
The reason for this superior performance lies in the known fact that even the very highest lying excited pseudo-states can make a significant contribution to the Fermi contact term.
And in the Lanczos algorithm it is precisely these pseudo-states, for which good approximations are obtained first.

One disadvantage of the Lanczos algorithm is however that the iterations have to be started with one of the property gradient vectors and our investigation showed that the convergence pattern strongly depends on which property gradient is chosen.
The best results are obtained, when it is the Fermi contact property gradient of one of the two coupled atoms.
This implies that one unfortunately has to carry out the Lanczos iterations for each coupling constant separately. On the other hand, this 
allows for a trivial parallelization of these calculations.

In future studies it will thus be worthwhile to investigate, whether the Lanczos algorithm could equally well be applied to the calculation of the other two linear response contributions to the coupling constants, i.e. the spin-dipolar and paramagnetic spin-orbit contributions.

\section*{Acknowledgements}

\section*{DATA AVAILABILITY}
The data that support the findings of this study are available
from the corresponding author upon reasonable request.

\section*{References}
%

 \end{document}